\documentclass{elsarticle}

\usepackage{amsmath}
\usepackage[colorlinks,citecolor=blue,urlcolor=blue,filecolor=blue,backref=page]{hyperref}
\usepackage{psfrag,epsf}
\usepackage{enumerate}
\usepackage{url}
\usepackage[utf8]{inputenc}
\usepackage{amssymb}
\usepackage{algorithm}
\usepackage{bbm}
\usepackage{algpseudocode}
\usepackage{pifont}
\usepackage{cancel}
\usepackage{caption}
\usepackage{xcolor}
\usepackage{multirow}
\usepackage{arydshln}
\usepackage{subcaption}
\usepackage{appendix}

\journal{Computational Statistics \& Data Analysis}

\begin{document}

\begin{frontmatter}



\title{Improving performances of MCMC for Nearest Neighbor Gaussian Process models with full data augmentation}

\author[pau]{Sébastien Coube-Sisqueille\footnote{sebastien.coube@univ-pau.fr}~}
\author[pau,sid]{Benoît Liquet\footnote{benoit.liquet@univ-pau.fr}~}
\address[pau]{ 
Laboratoire de Mathématiques et de leurs Applications, Université de Pau et des Pays de l'Adour, UMR CNRS 5142, E2S-UPPA, Pau, France\\ 
}
\address[sid]{
Department of Mathematics and Statistics, Macquarie University, Sydney\\

\vspace{5pt}
Funding was provided by the Energy Environment Solutions (E2S-UPPA) consortium and the BIGCEES project from E2S-UPPA (``Big model and Big data in Computational Ecology and Environmental Sciences")
}
\begin{abstract}
Even though Nearest Neighbor Gaussian Processes (NNGP) 
alleviate considerably MCMC implementation of Bayesian space-time models,  they do not solve the convergence problems caused by high model dimension. Frugal alternatives such as response or collapsed algorithms are an answer.
An alternative approach is to keep full data augmentation but to try and make it more efficient. Two strategies are presented. \\
The first scheme is to pay a particular attention to the seemingly trivial fixed effects of the model. 
Empirical exploration shows that re-centering the latent field on the intercept critically improves chain behavior. 
Theoretical elements support those observations.
Besides the intercept, other fixed effects may have trouble mixing. This problem is addressed through interweaving, a simple method that requires no tuning while remaining affordable thanks to the sparsity of NNGPs. \\
The second scheme accelerates the sampling of the random field using Chromatic samplers. This method makes long sequential simulation boil down to group-parallelized or group-vectorized sampling. The attractive possibility to parallelize NNGP density can therefore be carried over to field sampling.   \\
A \textsf{R} implementation of the two methods for Gaussian fields is presented in the public repository \\ \url{https://github.com/SebastienCoube/Improving_NNGP_full_augmentation}. An extensive vignette is provided. The presented implementation is run on two synthetic toy examples along with the state of the art package \textsf{spNNGP}. Finally, the methods are applied on a real data set of lead contamination in the United States of America mainland.
\end{abstract}
\end{frontmatter}
\section{Introduction}
\label{sec:intro}
Many social or natural phenomena happen at the scale of a territory and must be observed at various sites and possibly times. The rise of modern GPS and Geographic Information Systems made large and high-quality point-referenced data sets more and more available. Assume that, in a collections of sites  $\mathcal{S}$ of the space or space-time domain $\mathcal{D}$,  we have measurements $z(\cdot)$ with some kind of space or space-time  coherence. This coherence can be accounted for by introducing a spatially-indexed process $w(\cdot)$ that has a well-defined joint distribution on any finite subset of the domain. 
We consider a Gaussian model where the observations $z(\cdot)$ have been perturbed by a Gaussian noise $\epsilon$ of standard deviation $\tau$. Many models also add linear regression on covariates $X(\cdot)$, giving the following classical model formulation 
\begin{equation}
\label{equation:space_time_model}    
z(s) = \beta_0 + X(s)\beta^T+w(s)+\epsilon(s), s\in \mathcal{S}.
\end{equation}
In order to keep notations shorter, for any collection of spatial locations $\mathcal{P}\subset\mathcal{S}$, we denote the vector $\{w(s) : s\in \mathcal{P}\}$ as $w(\mathcal{P})$.
Gaussian processes (GP) make an elegant prior distribution for $w(\cdot)$  for continuous data, see \cite{Handbook_Spatial_Stats}. 
The GP prior distribution of $w(\mathcal{S})$ is $\mathcal{N}(\mu, \Sigma)$. The mean parameter of $w(\cdot)$ is usually fixed to $\mu = 0$ to avoid identification problems with the linear regression intercept $\beta_0$.
The covariance matrix is computed using a positive definite function $k(\cdot)$ with covariance parameters $\theta$, such as Matérn's covariance and its exponential and squared-exponential special cases. 
It can then be written as $\Sigma(\mathcal{S}, \theta)$, and its entries are $\Sigma(\mathcal{S}, \theta)_{i,j} = k(s_i, s_j, \theta)$. We denote $f(\cdot|\mu, \Sigma)$ the GP density, and we abbreviate it as $f(\cdot|\mu, \theta)$ .
The covariance parameters can have modeller-specified hyperpriors developed in \cite{pc_prior_fuglstad2015interpretable, NNGP}.

The weakness of GPs is that computing the GP prior density of $w(\mathcal{S})$ involves the determinant and inverse of $\Sigma(\mathcal{S}, \theta)$, incurring a computational cost that is cubic in the size of $\mathcal{S}$. Vecchia's approximation to Gaussian likelihoods received increased attention the past years, with theoretical developments  of \cite{General_Framework, Guinness_permutation_grouping, NNGP, finley2019efficient} and software presented in \cite{GpGp, finley2017spnngp}. The
Nearest Neighbor Gaussian Process (NNGP) is a special case of Vecchia's approximation that provides a surrogate of the inverse Cholesky factor of $\Sigma$ and uses it to approximate GP prior density. It starts by finding an ordering for the $n$ locations of $\mathcal{S}$ which we will denote $(s_1, \ldots ,s_n)$.
The ordering may have an impact on the quality of the approximation, and is discussed in \cite{NNGP, Guinness_permutation_grouping}. The joint latent density of $w(s_1,\ldots,s_n)$  is then written under the recursive conditional form
$$
f(w(s_1,\ldots, s_n)|\mu,\theta) =  f(w(s_1)|\mu,\theta)  \times\Pi_{i=2}^nf(w(s_i)|w(s_1,\ldots,s_{i-1}),\mu,\theta).
$$
Since $f(w(s_1,\ldots, s_n)|\mu,\theta))$ is a Multi-Variate Normal (MVN) distribution function, the conditional density 
$f(w(s_i)|w(s_1,\ldots,s_{i-1}),\mu,\theta), i\in 2 ,\ldots, n$ is a Normal as well. 
A NNGP is obtained by replacing the vector $w(s_1,\ldots, s_{i-1})$ that conditions $w(s_i)$ by a much smaller parent subset denoted  $w(pa(s_i))$ for each conditional density. The NNGP approximation to the GP prior joint density of $w(\cdot)$ is defined as 
\begin{equation}
\label{NNGP_formula}
\tilde f(w(s_1,\ldots, s_n)|\mu,\theta) = f(w(s_1)|\mu,\theta)\times
\Pi_{i=2}^n f(w(s_i)|w(pa(s_i)),\mu,\theta).
\end{equation}
This very general principle can be applied to any kind of well-defined multivariate density. However, as far as we know, MVN density approximation is the only application. This may be explained by the fact that non-Gaussian data can be handled with GP modelling thanks to link functions. Moreover, a NNGP defines a MVN density and it is possible compute explicitly and easily the sparse Cholesky factor of the precision matrix. 
The choice of the parents is critical but no universal criterion exists. A popular choice is to choose the parent locations $pa(s_i)$ as $s_i$'s nearest neighbors among $(s_1, \ldots ,s_{i-1})$, explaining the denomination ``Nearest Neighbors Gaussian Process" given in \cite{NNGP}. However,  \cite{NNGP, stein2004approximating} argue that mixing close and far-away observations can improve the approximation. 
This approximation is cheap and easily parallelisable. The latent density (\ref{NNGP_formula}) can be split into small jobs and dispatched to a cluster of calculators  (\cite{NNGP}). Its cost is linear in the number of observations under the condition that the size of each parent set is bounded. More advanced strategies exist such as grouping, proposed by Guinness in  \cite{Guinness_permutation_grouping}.  

If NNGPs work around the bottleneck of GP likelihood computation, they do not solve the problem of slow MCMC convergence. 
In \cite{NNGP}, the Gibbs sampler loops over $\theta$, $w(\mathcal{S})$ and $\beta$, $\mu$ is fixed to $0$. The latent field $w(\mathcal{S})$ is updated sequentially or by blocks.
This sampler suffers from slow mixing, in particular when $n$ increases. 
Other strategies have been proposed by Finley and al. \cite{finley2019efficient} that precisely avoid to sample the field in order to reduce the dimension of the model.  Yet another method  \cite{finley2019efficient, zhang2019practical} is to use convenient conjugate distributions for models where the range of $w(\cdot)$ and the variance ratio of $w(\cdot)$ and $\epsilon(\cdot)$ is fixed, and select the fixed parameters by cross-validation.  
Our approach is nevertheless to improve implementations of NNGP models where the latent field is explicitly sampled. Our first reason is that there may be situations where some of the methods presented in \cite{finley2019efficient} perform poorly while full data augmentation works well. 
For example, the \textit{collapsed NNGP} of \cite{finley2019efficient} enjoys low dimensionality and allows nonetheless to retrieve the latent field, but demands Cholesky factorization of large sparse matrices which may be unfeasible depending on $n$ and the dimension of $\mathcal{D}$.
The \textit{Response NNGP} of \cite{finley2019efficient} retrieves the covariance parameters $\theta$ but not the latent field $w(\mathcal{S})$. 
Our second reason is that efficient Gibbs sampler architectures can  sharply improve mixing. A NNGP defines a Markov Random Field, allowing to use the blocking methods of \cite{knorr2002block}. The sparse Cholesky factor in a NNGP  makes it possible to use the Ancillary-Sufficient Interweaving Strategy (AS-IS) presented in \cite{yu2011center}. 
The third reason is that full latent field sampling is all terrain, and can address many data models or be plugged into complex, non-stationary models like \cite{heinonen2016non}, while collapsed MCMC or conjugate models are much pickier\color{
black}.

Here is an outline of the article. 
Section \ref{sec:centering} focuses on the seemingly trivial fixed effects of the hierarchical model. 
In \ref{subsection:two_formulations_same_model} we propose a mild but efficient centering of the latent field on the least squares regression intercept. 
In \ref{subsection:Extension_to_other_fixed_effects}, we extend centering to other fixed effects, and we use interweaving from \cite{yu2011center} to propose a robust, tuning-less application.
Section \ref{sec:chromatic} targets the simulation of the random field. In \ref{subsection:Chromatic_samplers}, we propose to use the chromatic samplers developed by Gonzalez and al. in \cite{Gonzalez_parallel_gibbs} in order to carry the attractive parallelizability of NNGP density over to field sampling.  In \ref{subsection:chromatic_experiments}, we analyze the sensitivity of NNGP graph coloring and we benchmark coloring algorithms.  
We apply our methods in section \ref{sec:implementation}. We present our implementation (available at \url{https://github.com/SebastienCoube/Improving_NNGP_full_augmentation}) in \ref{subsection:about_our_implementation}. We test our implementation along with the state of the art package \textsf{spNNGP} presented in \cite{finley2017spnngp} on synthetic toy examples in \ref{subsection:toy_examples}. 
In \ref{subsection:lead_contamination}, we present an application on lead contamination  in the mainland of the United States of America. 
The article ends by a discussion in Section \ref{sec:discussion}.

\section{Latent field centering}
\label{sec:centering}

\subsection{Centering the latent field on the intercept}
\label{subsection:two_formulations_same_model}
\begin{figure}[]
\begin{subfigure}{.45\textwidth}
  \centering
  \includegraphics[width=.99\linewidth]{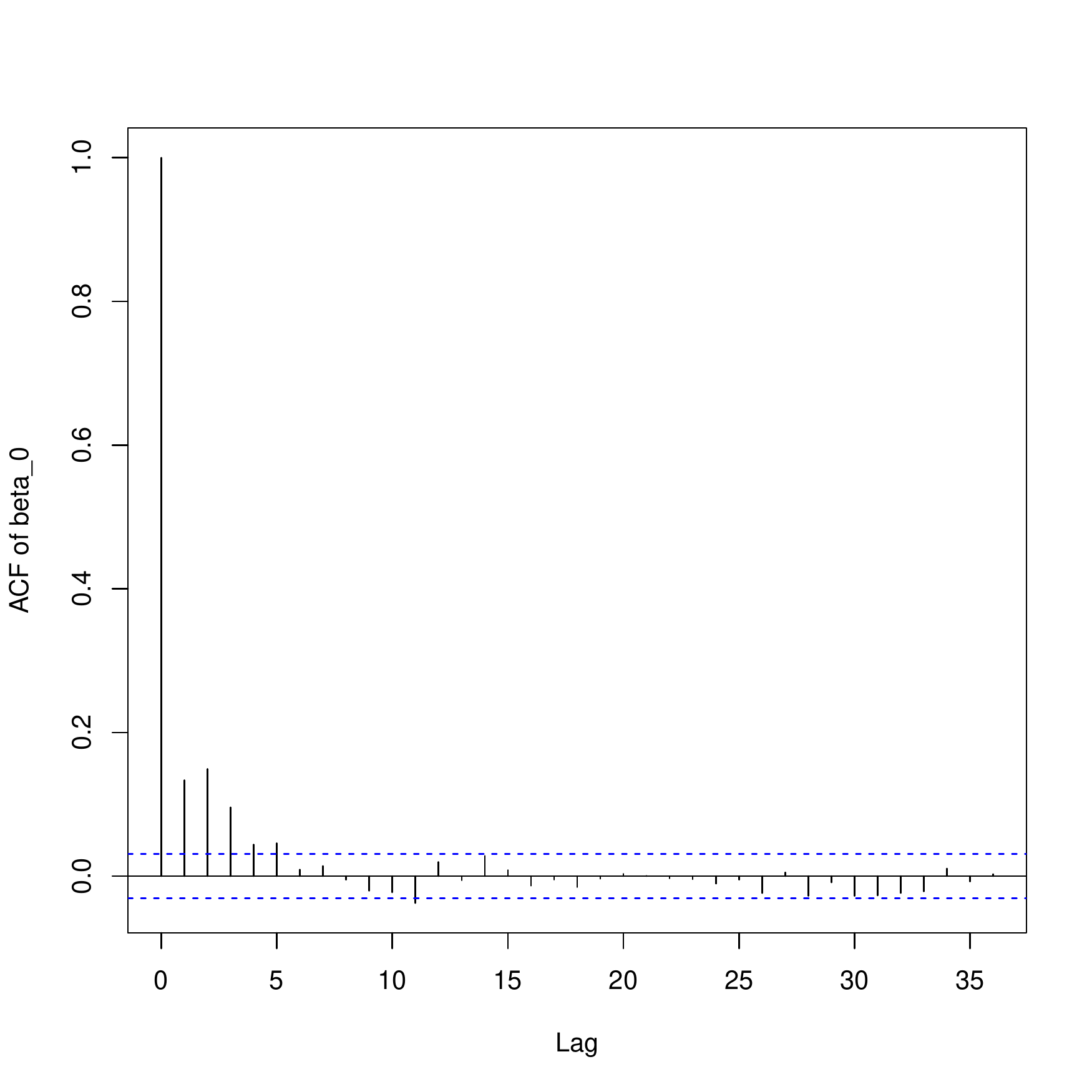}  
  \caption{\textsf{Centered model}}
  \label{fig:ACF_centered}
\end{subfigure}
\begin{subfigure}{.45\textwidth}
  \centering
  \includegraphics[width=.99\linewidth]{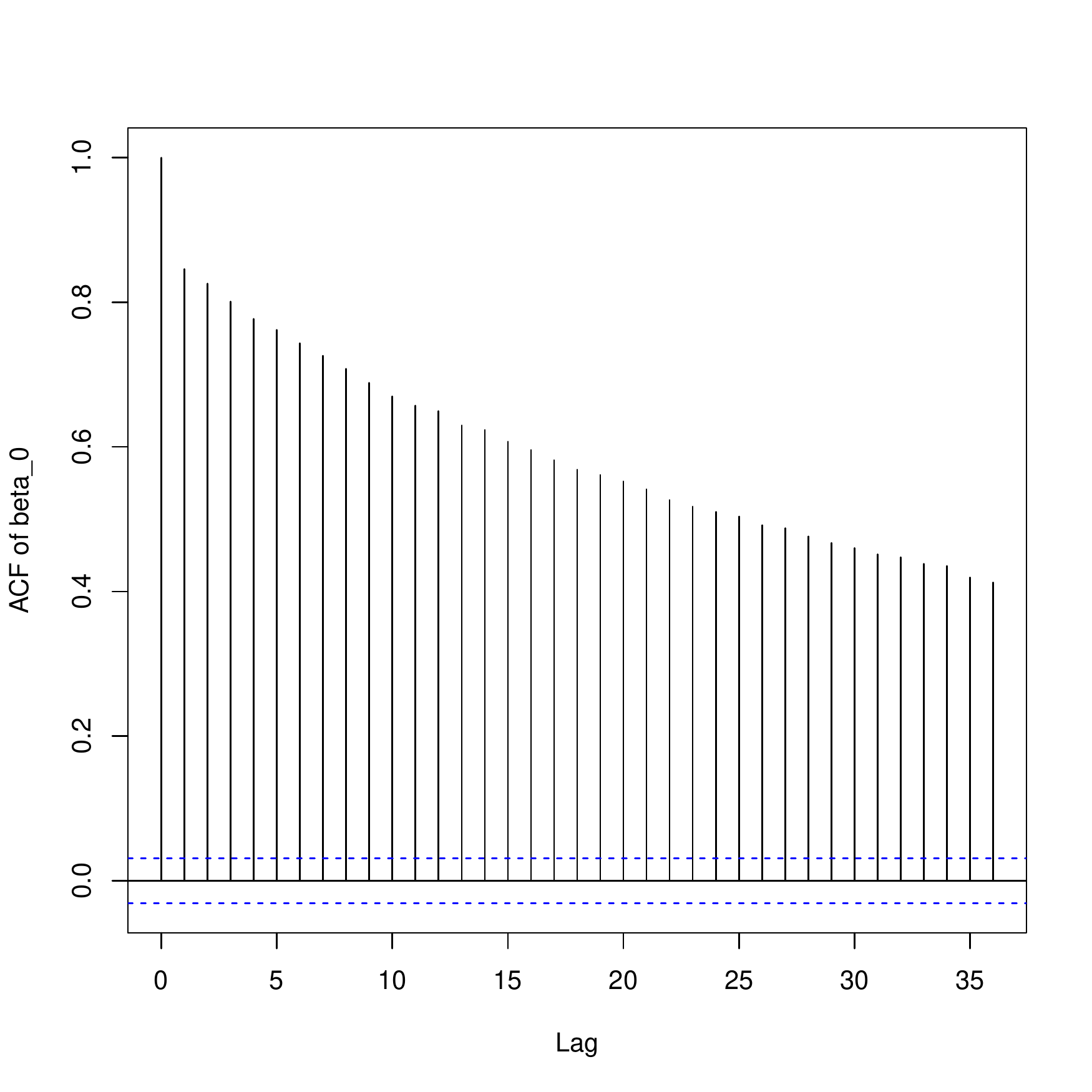}  
  \caption{\textsf{Standard model}}
  \label{fig:ACF_uncentered}
\end{subfigure}
\caption{The ACF of $\beta_0$ drops much faster in the centered model than in the standard model}
\label{fig:ACF}
\end{figure}

\begin{figure}[]
\begin{subfigure}{.45\textwidth}
  \centering
  \includegraphics[width=.99\linewidth]{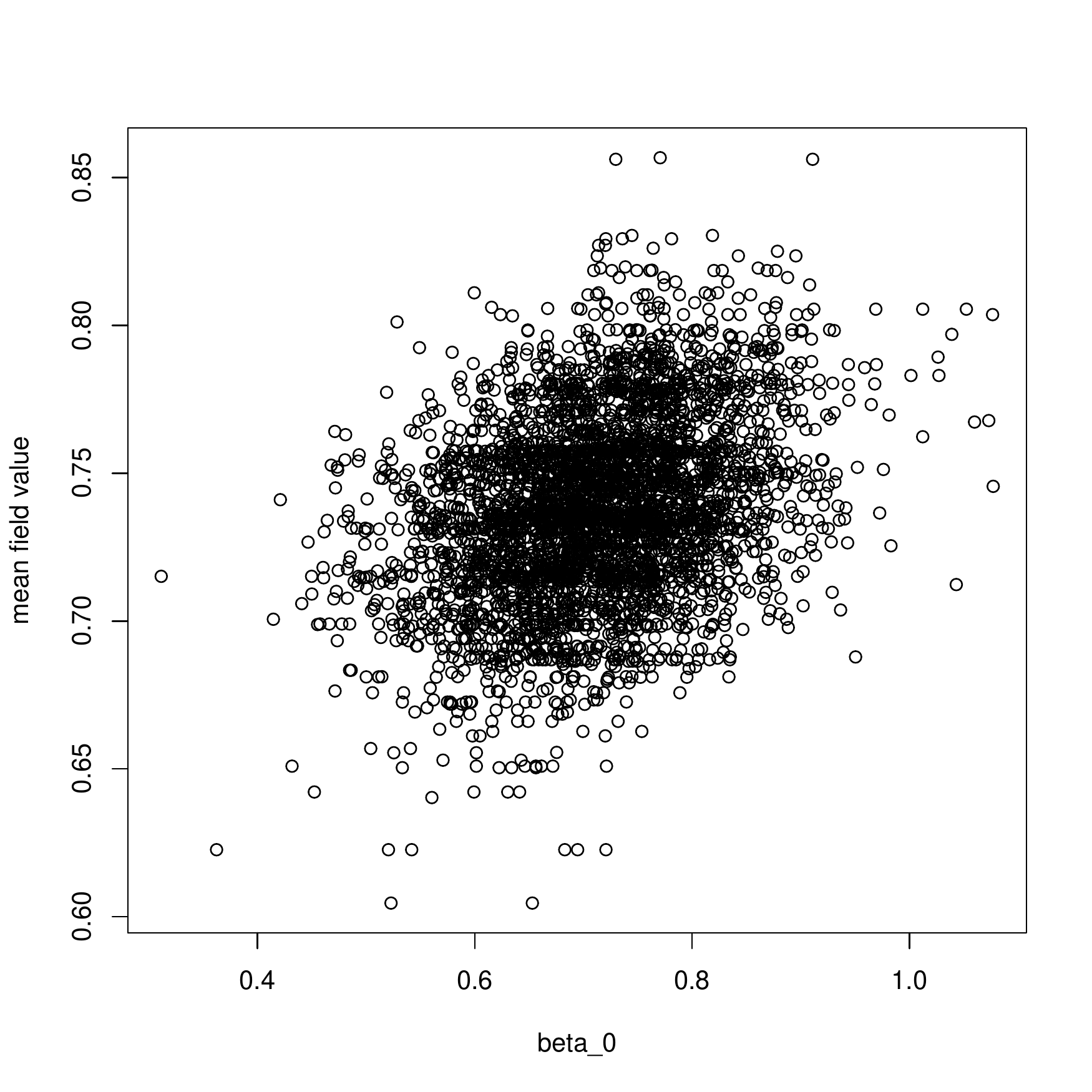}  
  \caption{\textsf{Centered model}}
  \label{fig:scatter_centered}
\end{subfigure}
\begin{subfigure}{.45\textwidth}
  \centering
  \includegraphics[width=.99\linewidth]{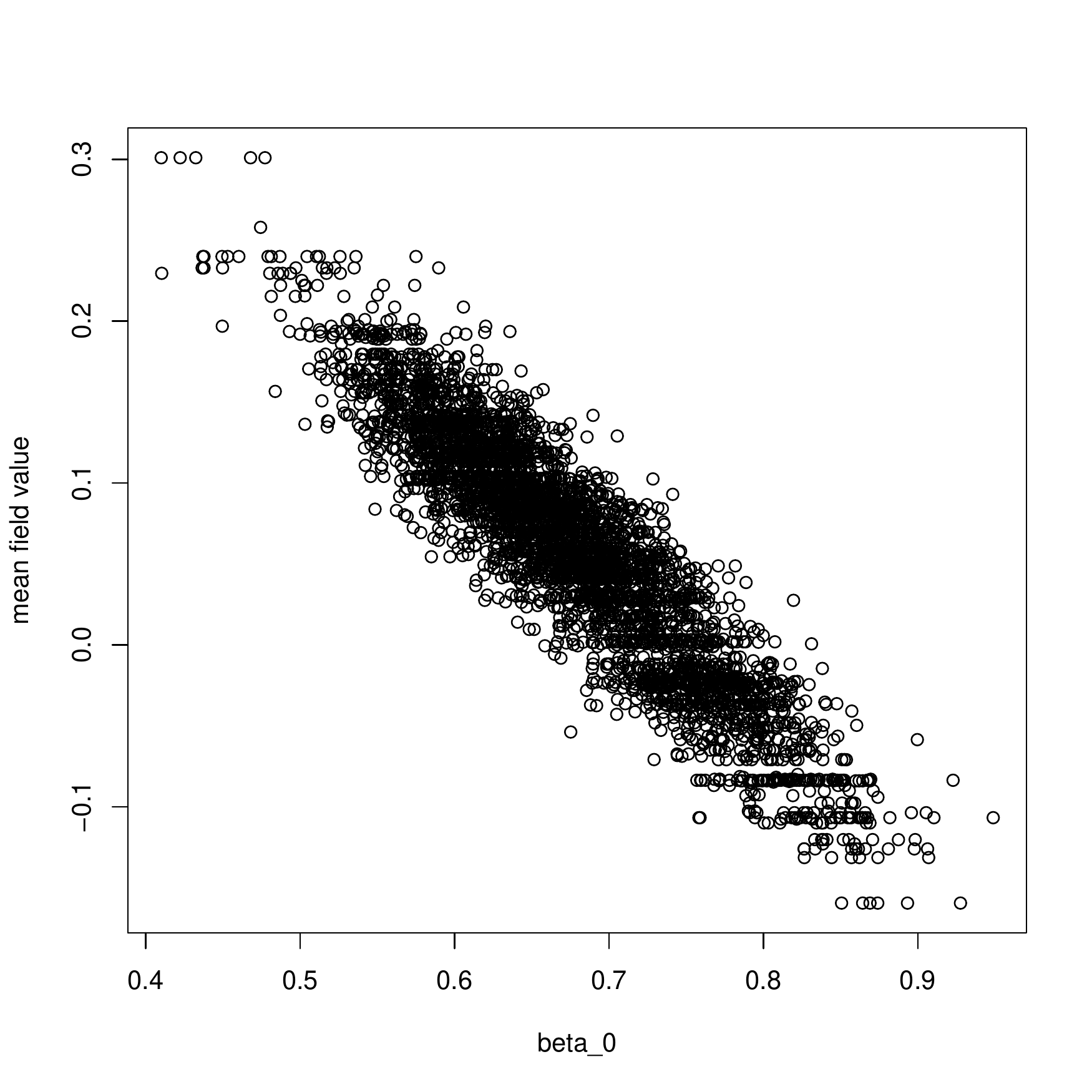}  
  \caption{\textsf{Standard model}}
  \label{fig:scatter_uncentered}
\end{subfigure}
\caption{When plotting $\frac{1}{n} \Sigma w(\mathcal{S})$ against $\beta_0$, the standard model exhibits a ridge-shaped point cloud}
\label{fig:scatter}
\end{figure}

The mean parameter $\mu$ of the prior density for the latent field $w(\cdot)$ is usually set to  $0$ in order to avoid identification problems with the intercept $\beta_0$. We call this formulation standard, since it is found in state of the art papers such as \cite{NNGP, finley2019efficient}. We name samples of the standard formulation $w_s(\cdot)$.  
Our proposal is to replace $w_{s}(\mathcal{S})$ by  a centered $ w_c(\mathcal{S}) =  w_s(\mathcal{S}) + \beta_0$ in the Gibbs Sampler. This substitution is a non degenerate linear transform that keeps the model valid, while keeping the possibility to transform the  samples back to standard parametrization if needed.
The centered parametrization can also be seen as a slightly different model, with \eqref{equation:space_time_model} becoming
\begin{equation}
\label{equation:space_time_model_centered}
z(s) = X(s)\beta^T+w_c(s)+\epsilon(s), s\in \mathcal{S}
\end{equation}
and the prior density of $w_c(\mathcal{S})$ becoming 
$$\tilde f(w_c(\mathcal{S})|\mu = \beta_0,\theta).$$ 
Those changes impact the full conditional distributions. Table \ref{tab:full_conditional} summarizes the changes in a Gibbs sampler for a Gaussian model found in \cite{NNGP}. 
We denote $f (\cdot |\cdot,\cdot)$ the normal density function, and  $\tilde Q$ the latent field's precision matrix defined by the NNGP. We abbreviate the interest variables $X(\mathcal{S})$ as $X$. We denote the vector made of $n$ times $1$ as $\textbf{1}$. The matrix obtained by adding $\textbf{1}$ to the left side of $X$ is named $[\textbf{1}|X]$. 
We did not feature prior distributions on the high-level parameters like $\theta$, $\tau$ or $\beta$ : their full conditionals would not be affected since centering changes only the NNGP prior and the observed data likelihood.
\begin{table*}
    \caption{Changes in the full conditional distributions}
    \centering
    \begin{tabular}{cll}
    Variable & Standard &  Centered  \\
    \hline
    $\beta_0$ 
    & 
    & $f (\beta_0, (\textbf{1}^T\tilde Q\textbf{1})^{-1}(\textbf{1}^T\tilde Q w_c) , 
    (\textbf{1}^T\tilde Q\textbf{1})^{-1})$
    \\
    
    $\beta$ 
    & 
    & $f (\beta, (X^T X)^{-1}(X^T (z-w_c))) , 
    \tau^2(X^T X)^{-1})$
    \\
    
    $(\beta_0, \beta)$
    &  $f (\beta, ([\textbf{1}|X]^T [\textbf{1}|X])^{-1}([\textbf{1}|X]^T (z-w_s))) , $
    &  \\
    
    & $ \tau^2([\textbf{1}|X]^T [\textbf{1}|X])^{-1}) $
    &  \\
    
    $\theta$ 
    & $ \tilde f (w_s(\mathcal{S})|0, \theta)$ 
    & $\tilde f (w_c(\mathcal{S})|\beta_0, \theta)$\\
    
    $\tau$ & 
    $ \Pi_{s\in \mathcal{S}} f (z(s)| w_s(s) + \beta_0 + X(s)\beta^T, \tau ) $  &
    $\Pi_{s\in \mathcal{S}} f (z(s)|w_c(s) + X\beta^T, \tau ) $ \\
    
    $w(x)$
    & $\tilde f(w_s(x)|w_s(\mathcal{S}\backslash x), 0, \theta)$ 
    &$\tilde f(w_c(x)|w_c(\mathcal{S}\backslash x), \beta_0, \theta) $ \\ 
    &$f(z(x)|w_s(x) + \beta_0 + X(x)\beta^T, \tau)$ & $f(z(x)| w_c (x) + X(x)\beta^T, \tau) $ \\
    \end{tabular}
    \label{tab:full_conditional}
\end{table*}{}
\noindent
Even if the modification is minor, the improvement in the mixing of the intercept is clear. We simulated a little toy example with $1000$ observations and we ran the two Gibbs samplers. The autocorrelation plots (Figure \ref{fig:ACF}) are clearly in favor of the centered formulation. Even though for this toy example the standard model mixes after a few hundred iterations, this is not the case for larger cases.\\
We observed empirically that there is much more correlation between $w(\mathcal{S})$ and $\beta_0$ in the standard implementation. Plotting $\frac{1}{n} \Sigma w(\mathcal{S})$ against $\beta_0$ (Figure \ref{fig:scatter}) displays a clear ridge in the case of the standard model (Figure \ref{fig:scatter_uncentered}). 
This means that the whole latent field has to shift upwards and downwards for the intercept to explore its posterior distribution, causing a slow exploration. \\
Ridge-like densities are a well known plague of Gibbs samplers, and linear recombination is one of the tools to get rid of it, see  \cite{gelfand1995efficient}.  


The behavior of the toy example arises from the fact that the fraction of $\beta_0^t$ that is carried over in $w^{t+1}$ and $\beta_0^{t+1}$ changes following the model. 
Take a simpler spatial model where only an intercept and the Gaussian latent field are estimated, while the Gaussian noise variance $\tau^2$ and the NNGP precision $\tilde Q$ are known. The intercept coefficient has an improper constant prior. Assume that the latent field is sampled in one step (which is usually not the case unless the data size is very small). \\
Denote the diagonalization $\tilde Q = V^T \lambda V$, $V$ being a square matrix of eigenvectors and $\lambda$ being a diagonal matrix of eigenvalues. The eigenvalues are positive since $\tilde Q = \tilde R^T\tilde R$. Note $\alpha_i$, $i = 1,\ldots,n$ the coordinates of the vector $\textbf{1} = (1,\ldots,1)$ in the orthonormal basis $V$. 
The following results are proved in \ref{section:stochastic_form_centering}. 
\\
The first result is that a fraction of $\beta_0$ is carried over in the empirical mean of the latent field. Note $\rho = \Sigma_{i=1}^n\alpha_i^2(\tau^2\lambda + I_n)^{-1}_{i, i}/n$.
Then, 
$$\overline{w_s^{t+1}} = \mu_s -\rho \beta_0^t + \epsilon_s^{t+1} ~~~ \mbox{and} ~~~ \overline{w_c^{t+1}} = \mu_c + (1-\rho) \beta_0^t + \epsilon_c^{t+1}, $$
$\mu_s$ and $\mu_c$ being fixed,  $\epsilon_s$ and $\epsilon_c$ being stochastic innovations, and $t$ being the index of the iteration. 
Moreover, we have $$0 \leq \rho \leq 1.$$
Thanks to this result, we can see that if a high fraction of $\beta_0$ is carried over in the mean of the standard latent field, then a low fraction of $\beta_0$ will be carried over in the mean of the centered latent field, and conversely.
This is clearly what can be seen in figure \ref{fig:scatter}.

The next question is why $\rho$ is closer to $1$ than to $0$.
This point is difficult to clarify because there is no analytic expression for terms where $\tilde Q$ is involved. 
For example, the range parameters are updated through a Metropolis step in \cite{NNGP} because a full conditional draw is challenging. 
However, we can start from $\rho = \Sigma_{i=1}^n\alpha_i^2(\tau^2\lambda + I_n)^{-1}_{i, i}/n$ and make a deduction : if the sum is high,  then $\lambda_{i, i}$ is small when $\alpha_i$ is big ; and conversely $\lambda_{i, i}$ is big when $\alpha_i$ is small.
We can re-formulate~: $\lambda^{-1}_{i, i}$ is big when $\alpha_i$ is big. 
Now, remark that $V$  and $\lambda^{-1}$  respectively are  the spatial basis and coefficients of the Karhunen-Loève decomposition of the NNGP prior.
This means that $\alpha_i$ is high for the first components of the decomposition, where $\lambda^{-1}$ is the highest. 
In other terms, $\textbf{1}$ ``resonates" with the first spatial basis functions of the Karhunen-Loève decomposition. This conclusion is consistent with the fact that $\tilde Q$ parametrizes a spatially coherent latent field, inducing that a few spatial basis functions are enough to describe most of $w$. For example, in the Predictive Process model of \cite{PP}, $w$ is approximated by a degenerate process with a low-rank covariance matrix. 

Now, let's focus on the expressions of $[\beta_0^{t+1}|\beta_0^t]$. Like the mean of the latent field, they can be expressed with a fixed part, a geometric carry-over, and an innovation. 
In the standard model, a fraction $\rho$ of $\beta_0^t$ is carried over. We already discussed this quantity. 
As for the centered model, the fraction of $\beta_0^t$ which is conserved in $\beta^{t+1}$ is 
$$ \Sigma_{i=1}^n\left((\alpha_i^2\lambda_{i, i})
(\tau^2\lambda_{i, i})/(\tau^2\lambda_{i,i}+1) \right)/\Sigma_{i=1}^n\alpha_i^2\lambda_{i, i}.
$$ 
Once again, the geometric term is between $0$ and $1$ since $0\leq (\tau^2\lambda_{i, i})/(\tau^2\lambda_{i,i}+1)\leq 1$.
Like before, suppose that $\alpha$ is big when $\lambda$ is small. Then, when  $\alpha_{i}$ is the largest, $(\tau^2\lambda_{i, i})/(\tau^2\lambda_{i,i}+1)$  will be much smaller than $1$, resulting in $\Sigma_{i=1}^n\left((\alpha_i^2\lambda_{i, i})
(\tau^2\lambda_{i, i})/(\tau^2\lambda_{i,i}+1) \right)$ being much smaller than $\Sigma_{i=1}^n\alpha_i^2\lambda_{i, i}$. Therefore, we can expect a small proportion of $\beta_0^t$ to remain in $\beta_0^{t+1}$.

\subsection{Adaptation to other fixed effects}
\label{subsection:Extension_to_other_fixed_effects}
Field centering can be extended to other fixed effects.
In most cases it is unnecessary because centering and scaling $X(\mathcal{S})$ is enough to considerably improve chain behavior.  Even worse, the Gibbs sampler usually behaves very bad if the random field is centered on other fixed effects than the intercept. There are nonetheless cases where bad mixing of the regression coefficients happens again. 
In this case, it is often useful to try and center $w(\cdot)$ not only on the intercept but also on the troublesome covariates' fixed effect. 
However, doing preliminary runs and picking manually which fixed effects the field needs to be centered on would be tedious.

Interweaving, introduced by Yu and Meng \cite{yu2011center}, combines the advantages of the two strategies and removes the need to choose. 

The method takes advantage of the discordance between two parametrizations to construct the following step~: 
$$ 
[Y_1|\theta^t] \rightarrow [Y_2|Y_1] \rightarrow[\theta^{t+1}|Y_2],
$$
$Y_1$ and $Y_2$ being two data augmentations and $\theta$ the parameter. 
Usually, it is complicated to sample directly $[Y_2|Y_1]$. Drawing an intermediary $\theta^{t+0.5}$ gives
$$ 
[Y_1|\theta^t] \rightarrow [\theta^{t+0.5}|Y_1] \rightarrow [Y_2|\theta^{t+0.5}, Y_1]  \rightarrow[\theta^{t+1}|Y_2].
$$
It is possible that $[Y_2|\theta, Y_1]$ is a deterministic transformation, giving a degenerate joint distribution. 
Note that interweaving is not alternating : an alternating scheme would be
$ 
[Y_1|\theta^t] \rightarrow [\theta^{t+0.5}|Y_1] \rightarrow  [Y_2|\theta^{t+0.5}] \rightarrow[\theta^{t+1}|Y_2].
$
The strategy is usually very efficient if the two parametrizations are an ancillary-sufficient couple, giving an Ancillary-Sufficient Interweaving Strategy (ASIS), and can even be efficient when none of the two augmentations performs well when implemented separately.

Algorithm \ref{algorithm:regression_updating_interweaving} presents the  steps to update the regression coefficients with interweaving.

The two parameterizations are $w$ which is un-centered, and $v$, which is centered on all the fixed effects. 

For the sake of simplicity, we suppose that there is only one measurement of  $X$ per spatial location and we use an improper constant joint hyperprior on $(\beta_0,\beta)$. The parameters that depend on the state in the Gibbs sampler are indexed by $t$.
If the observations were not Gaussian, step \textsf{4} would be left unchanged while step \textsf{2} would be adapted just like in any generalized NNGP model \cite{NNGP}.

There are two limitations to this approach. The first is the case where several measurements of the interest variable $z(\cdot)$ and the regressors $X(\cdot)$ are done at the same spatial location. The model must be extended as 
$$z(s, i) = X(s, i)\beta^T+w(s)+\epsilon(s, i), s\in \mathcal{S}, 1\leq i \leq m(s),$$
$m(s)\geq 1$ being the number of observations in the site $s$. 
In this setting, some variables vary within one spatial locations while other do not.
For example, the presence of asbestos in buildings may be considered as a location-wise regressor while smoking is an observation-wise regressor. 
If the regressors vary within one location, 
it is impossible to center the field on the corresponding fixed effects.  
This would mean that the normal random variable $w(s)$ has several mean parameters at the same time. 
However, it is still possible to restrict interweaving to the regression coefficients associated to the location-wise variables. Our implementation allows one to specify which regressors are associated to spatial location and which are associated to individual measurements. 
A NNGP being a point-measurement model, regressors obtained through gridded and areal data are immediately eligible for this method.

The second limitation is the computational cost. With improper constant prior, the centered regression coefficients follow a  MVN distribution whose mean and variance need to be computed at each update of $\theta$.  The sparsity induced by Vecchia's approximation is critical for the feasibility of the method because it ensures that matrix multiplications involving $\tilde Q$ are affordable. 
Using a sparse matrix formulation for $X$ could further alleviate this operation if $X$ has dummy variables or null measurements. \\
\begin{algorithm*}[t]
\caption{Regression coefficient updating with interweaving}
\label{algorithm:regression_updating_interweaving}
\begin{algorithmic}[1]
\State \textbf{input} 
$\tilde Q ^t, w^t, X, \beta^t, \beta_ 0^t, \tau^t $
\State   \textbf{simulate} $\beta_0^{t+.5}, \beta^{t+.5}$ following $\mathcal{N} (([\textbf{1}|X]^T [\textbf{1}|X])^{-1}([\textbf{1}|X]^T z) , \tau^2([\textbf{1}|X]^T [\textbf{1}|X])^{-1})$
\State   $v = w^t+ X(\beta^{t+.5})^T$
\State   \textbf{simulate} $\beta_0^{t+1}, \beta^{t+1}$ following  $\mathcal{N} (([\textbf{1}|X]^T\tilde Q^t[\textbf{1}|X])^{-1}([\textbf{1}|X]^T\tilde Q^t v) , ([\textbf{1}|X]^T\tilde Q^t[\textbf{1}|X])^{-1})$
\State   $ w^{t+0.5} = v - X(\beta^{t+1})^T$
\State  \textbf{return}  $\beta_0^{t+1} \beta^{t+1}, w^{t+0.5}$
\end{algorithmic}
\end{algorithm*}
\noindent
\section{Chromatic sampler for Nearest Neighbor Gaussian Process}\label{sec:chromatic}
\subsection{Chromatic samplers and how to apply them to NNGP}\label{subsection:Chromatic_samplers}
In a Gibbs sampler, the parameters of a model are updated sequentially.  If a set of variables happens to be mutually independent conditionally on the other variables of the model and are updated consecutively by the Gibbs sampler, their sampling can be parallelized. 
Let's consider a Gibbs sampler or a Metropolis-Within-Gibbs aiming to sample from a joint multivariate distribution $f(x_1, \ldots ,x_n)$.
$$
\begin{array}{rl}
     x_1^{t+1}&\sim f(x_1|x_2^t, \ldots ,x_n^t)  \\
      \ldots &\\ 
     x_i^{t+1}&\sim f(x_i|x_1^{t+1}, \ldots ,x_{i-1}^{t+1}, x_{i+1}^t ,\ldots, x_n^t)  \\
      \ldots &\\ 
     x_n^{t+1}&\sim f(x_n|x_1^{t+1} ,\ldots, x_{n-1}^{t+1}).  \\
\end{array}
$$
Let's introduce $p\leq n $ vectors $X_1 ,\ldots, X_p$ so that $(x_1 ,\ldots, x_n) = (X_1 ,\ldots, X_p)$, and suppose that  $\forall X \in X_1 ,\ldots, X_p$, either $X$ has only one element or the elements of $X$ are conditionally independent given the other variables. 
The Gibbs sampler can then be re-written \\
$$
\begin{array}{rl}
     x^{t+1}_i\in X_1 &\sim f(x_i|X_2^{t} ,\ldots, X_p^{t})  \\
      \ldots &\\ 
     x^{t+1}_i\in X_j &\sim f(x_i|X_1^{t+1} ,\ldots, X_{j-1}^{t+1},X_{j+1}^{t} ,\ldots, X_p^{t})  \\
      \ldots &\\ 
     x^{t+1}_i\in X_p &\sim f(x_i|X_1^{t+1} ,\ldots, X_{p-1}^{t+1}).  \\
\end{array}
$$
Since all elements from $X_j$ are simulated from independent densities, it is possible to parallelize their sampling. 

\begin{figure*}
\begin{subfigure}{.32\textwidth}
  \centering
  \includegraphics[width=.99\linewidth]{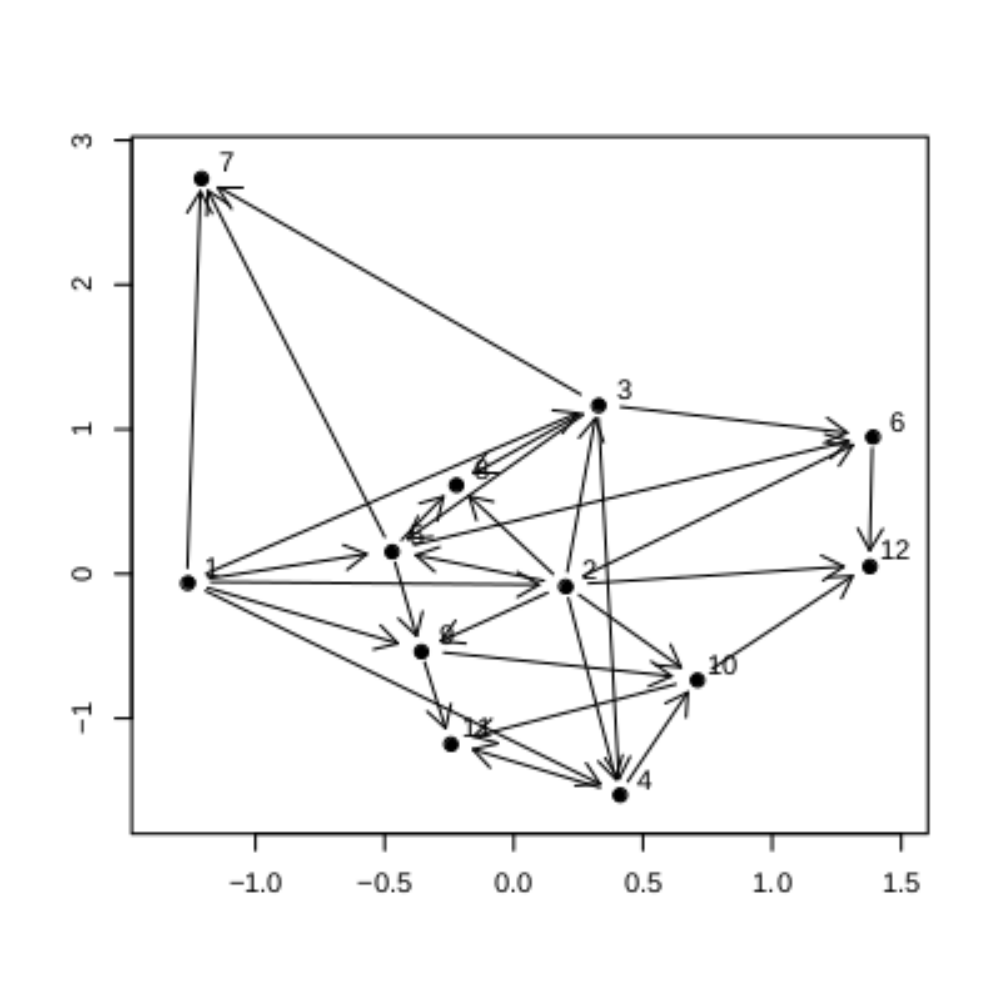}  
  \vspace{-2\baselineskip}
  \caption{\textsf{Directed Acyclic Graph}}
  \label{fig:DAG}
\end{subfigure}
\begin{subfigure}{.32\textwidth}
  \centering
  \includegraphics[width=.99\linewidth]{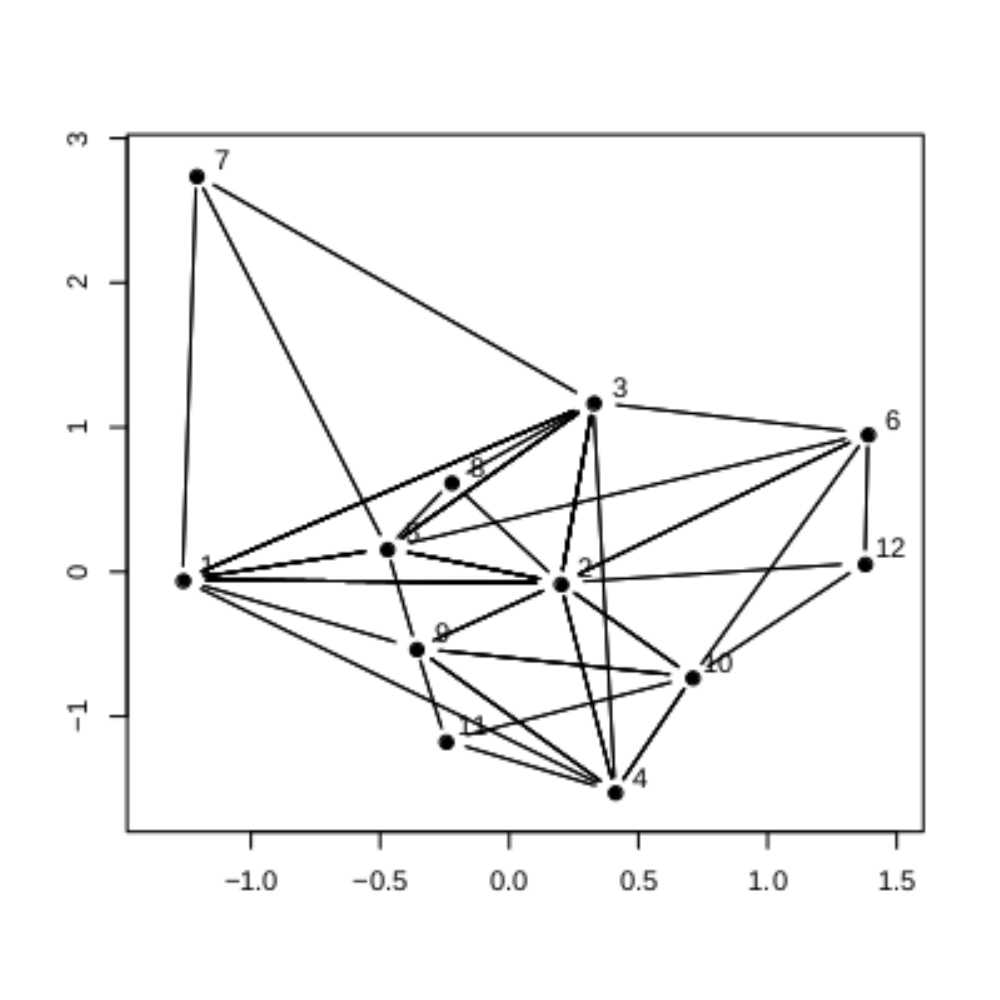}  
  \vspace{-2\baselineskip}
  \caption{\textsf{Moral Graph}}
  \label{fig:Moral}
\end{subfigure}
\begin{subfigure}{.32\textwidth}
  \centering
  \includegraphics[width=.99\linewidth]{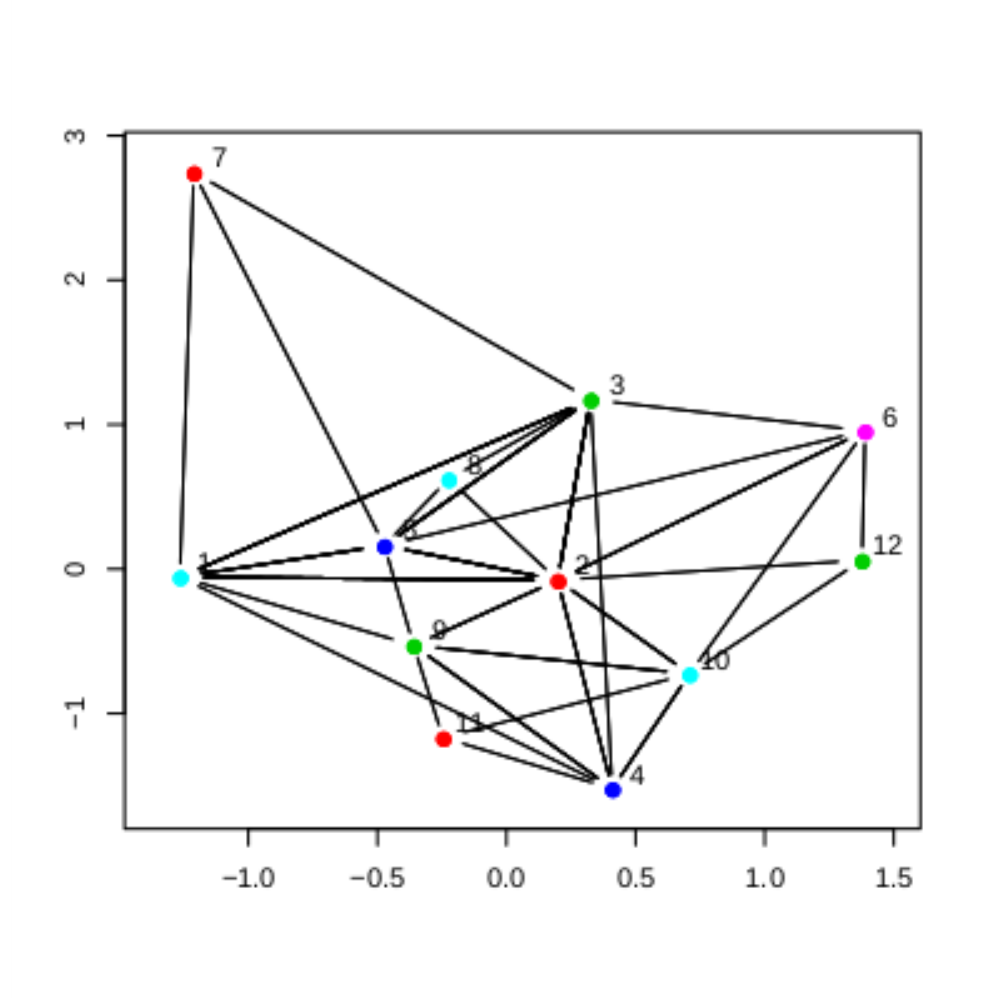}  
  \vspace{-2\baselineskip}
  \caption{\textsf{Colored Moral Graph}}
  \label{fig:Moral_colored}
\end{subfigure}
\caption{Moralization and coloring of a DAG}
\label{fig:coloring}
\end{figure*}

A NNGP  is defined on a Directed Acyclic Graph (DAG) by Datta and al. \cite{NNGP}, see \cite{General_Framework} for discussion about other Vecchia approximations. Then, using the argument of recursive kernel factorization given in \cite{graphical_models_Lauritzen}, it has the Markov properties on the moral graph obtained by un-directing the edges and ``marrying'' the parents in the DAG (Figure \ref{fig:coloring}).
Graph vertex coloring associates one color to each node of a graph while forbidding that two connected nodes have the same color, just like coloring a map while forbidding that two countries that share a border have the same color. 
Using inductively the Global Markov property, it is possible to  guarantee mutual conditional independence for the variables or the blocks that correspond to vertices sharing the same color.
Chromatic sampling can be applied straightforwardly to the Gibbs sampler presented in \cite{NNGP}. It also allows to compute normalizing constants and can be combined with the covariance parameter blocking proposed by Knorr-Held and Rue in \cite{knorr2002block}. 

Chromatic samplers can be applied to blocked sampling as well.
This method consists in updating the latent field in various locations at once.
Chromatic sampling is a special case of blocked sampling, because in general there is no conditional independence within one block.  
Precisely, sampling the latent field jointly in a region of the domain reduces the negative impact of spatial auto-correlation on the behavior of MCMC chains. 
Blocked sampling may be applied to the latent field alone \cite{NNGP} or improve both covariance parameters and field sampling in \cite{knorr2002block}. 
Even though there is no conditional independence within one block, there is some conditional independence between the blocks as long as there is no edge between any pair of their respective vertices, allowing for chromatic sampling.  
The matrix $B^TAB$ indicates the connections between the blocks, $A$ being the adjacency matrix of the NNGP latent field's Markov graph, and $B$ a vertex-block indicator matrix ($B_{i, j}=1$ if vertex $i$ belongs to block $j$). 
$~~~~$

\noindent

\subsection{Coloring of NNGP moral graphs : sensitivity analysis and benchmark of the algorithms}\label{subsection:chromatic_experiments}
Coloring the moral graph $\mathcal{G}^m$ is a critical step in chromatic sampling and determines the attractiveness of the method with respect to the ``vanilla'' versions of the algorithms (one-site sequential sampling or blocked sampling with several blocks). We focus on two variables to summarize the efficiency of chromatic sampling : 
\begin{itemize}
    \item The number of colors : the smaller this number, the fewer the number of steps in the chromatic sampler.
    \item The time needed for coloring, that must be small with respect to the running time of the MCMC chains
\end{itemize}
This section has two objectives. The first is to test the sensitivity of those two interest variables to the  properties of $\mathcal{G}^m$ and the coloring algorithm using variance-based sensitivity analysis. The second objective is to benchmark various coloring algorithms and find a rule to choose the algorithm.  \\
We test various factors that may change the structure of $\mathcal{G}^m$ :
\begin{itemize}
    \item Size $n$
    \item Number of parents in the DAG $m$
    \item Spatial domain dimension $d$
    \item Ordering of the points
\end{itemize}
We also test 3 coloring algorithms, given in detail in \ref{subsection:coloring_algo} : 
\begin{itemize}
    \item Naive greedy coloring : coloring each vertex with the smallest available color.
    \item Degree greedy coloring  : reorder the vertices following their number of neighbors, and apply naive greedy coloring.
    \item  DSATUR heuristic : color the node that has the highest number of distinct colors among its neighbors (\textit{Degree of SATURation}), and break ties using the number of neighbors.
\end{itemize}
The full results of the experiments are given in \ref{subsection:coloring_exp}, and the sensitivity analyses are summarized in table \ref{table:sensitivity_coloring}.

\subsubsection{Pilot experiment}

\underline{Design}\\
The objective is to do preliminary sensitivity analysis and benchmark on small graphs. 
We test for the three coloring algorithms and for graphs with the following attributes, each case being replicated $10$ times : 
\begin{itemize}
    \item Graph size $n = 500, 1000, 2000$.
    \item Number of parents $m = 5, 10, 20$.
    \item Dimension $d = 2, 3$.
    \item Ordering following the first coordinate from  \cite{NNGP}, at random, or using MaxMin heuristic from \cite{Guinness_permutation_grouping}. 
\end{itemize}

\underline{Sensitivity}

\begin{figure*}
\begin{subfigure}{.49\textwidth}
  \centering
  \includegraphics[width=1\linewidth]{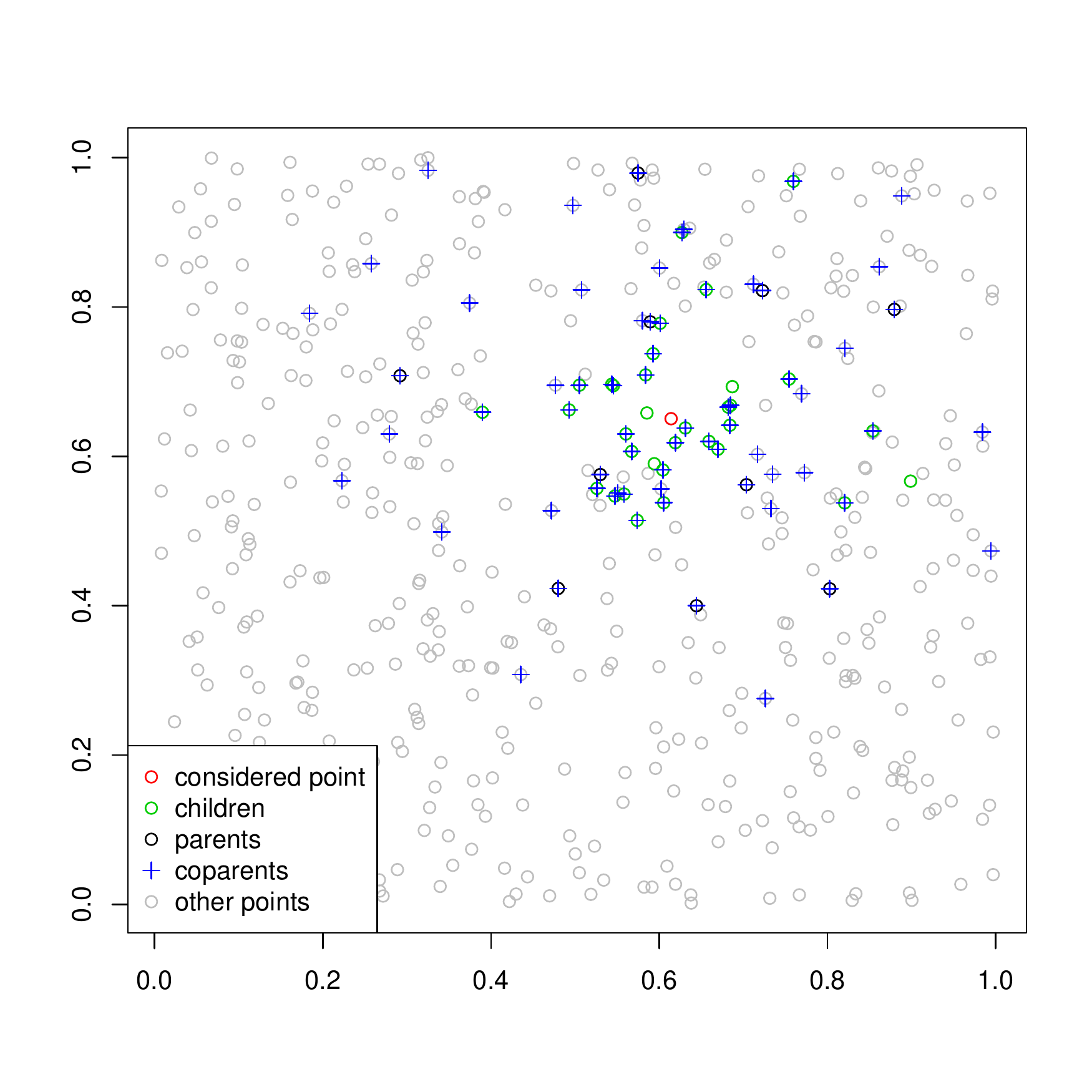}
  \caption{Max-min ordering \\(the considered point is the $30^{th}$ of $500$)}
  \label{fig:connexion_maxmin}
\end{subfigure}
\begin{subfigure}{.49\textwidth}
  \centering
  \includegraphics[width=1\linewidth]{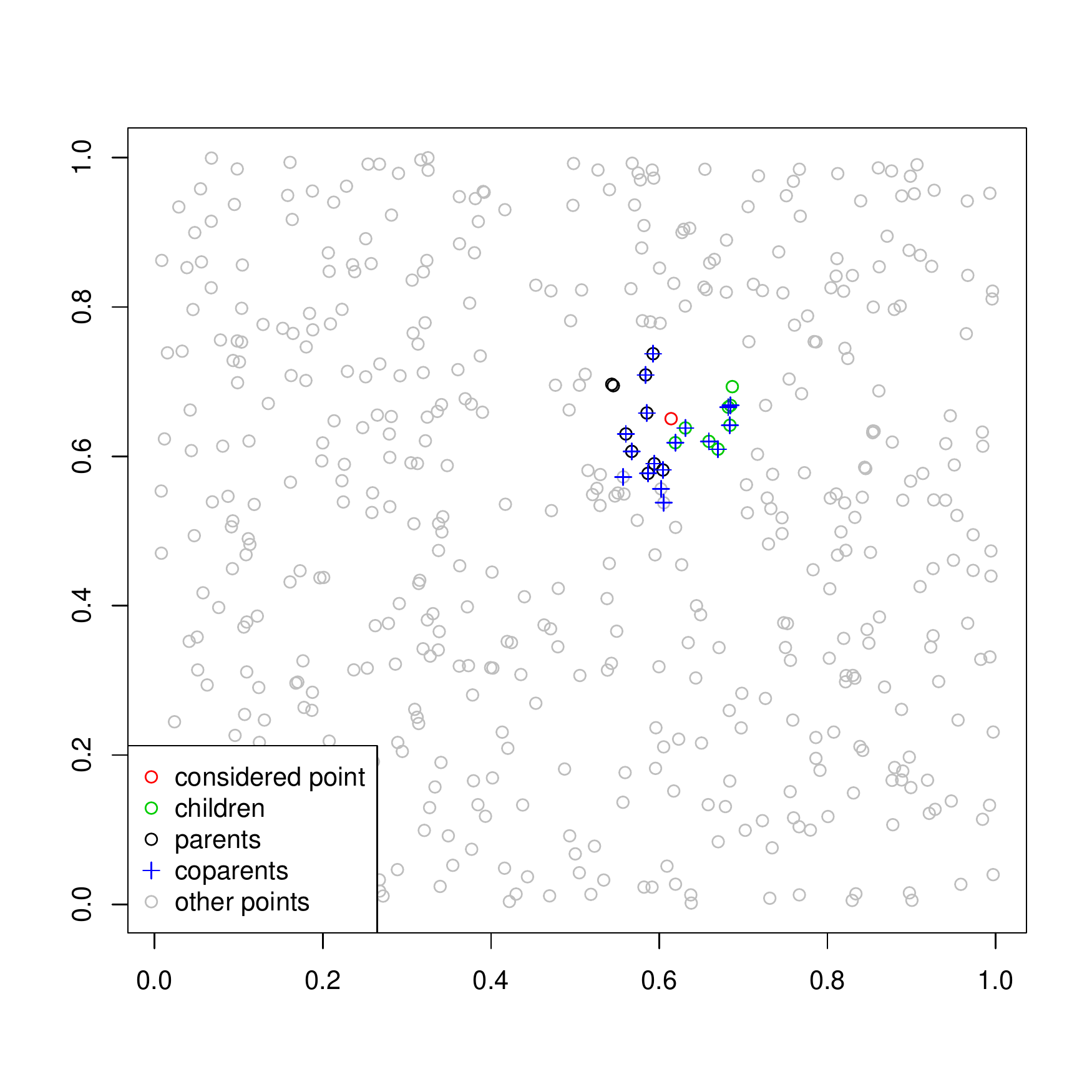}
  \caption{Coordinate ordering \\
  $~~~~~~$}
  \label{fig:connexion_coord}
\end{subfigure}
\caption{Connections of the same point, with two different orderings.}
\label{fig:markov_connexions}
\end{figure*}

\begin{figure*}
\begin{subfigure}{.49\textwidth}
  \centering
  \includegraphics[width=1\linewidth]{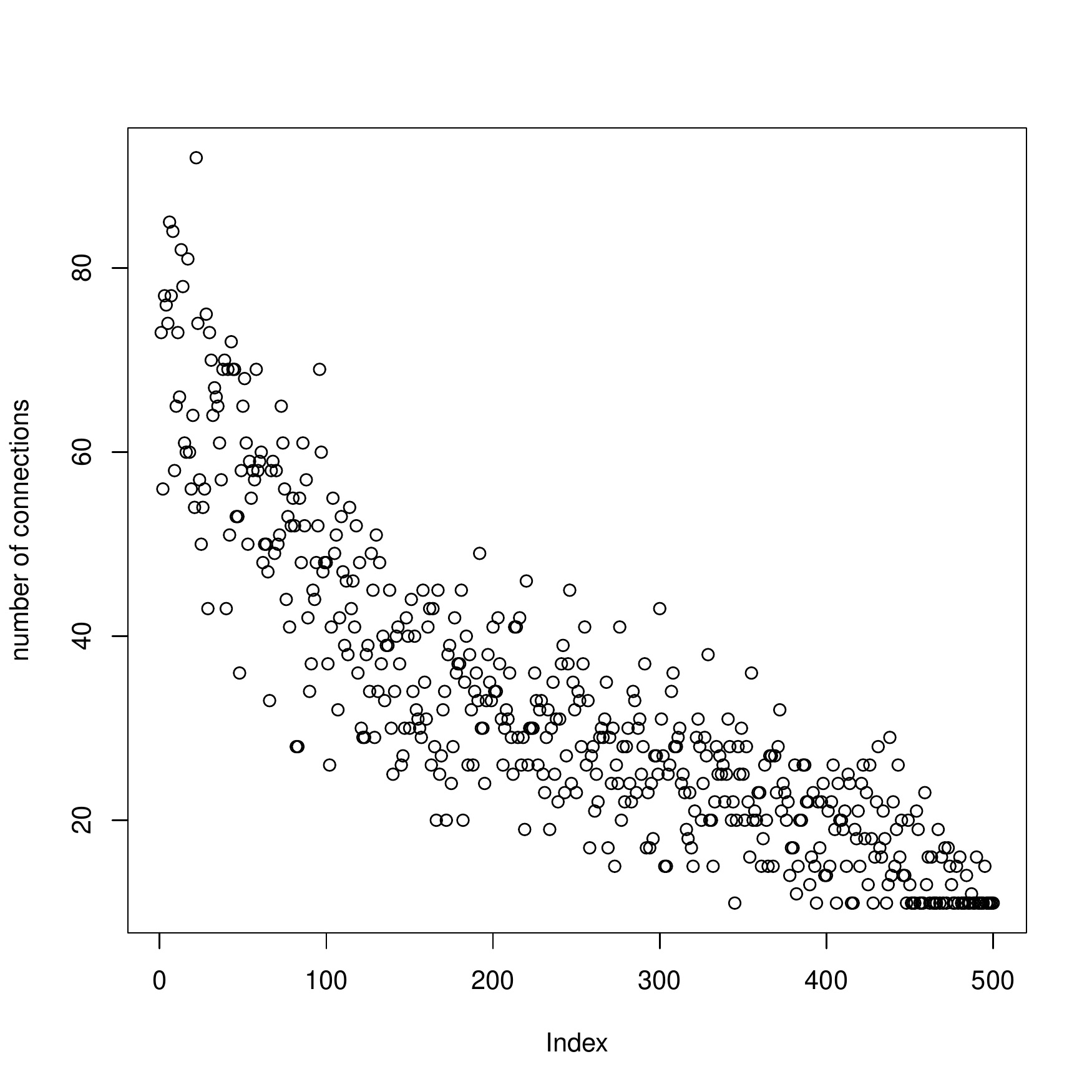}
  \caption{Max-min ordering}
  \label{fig:connexion_maxmin_ordered}
\end{subfigure}
\begin{subfigure}{.49\textwidth}
  \centering
  \includegraphics[width=1\linewidth]{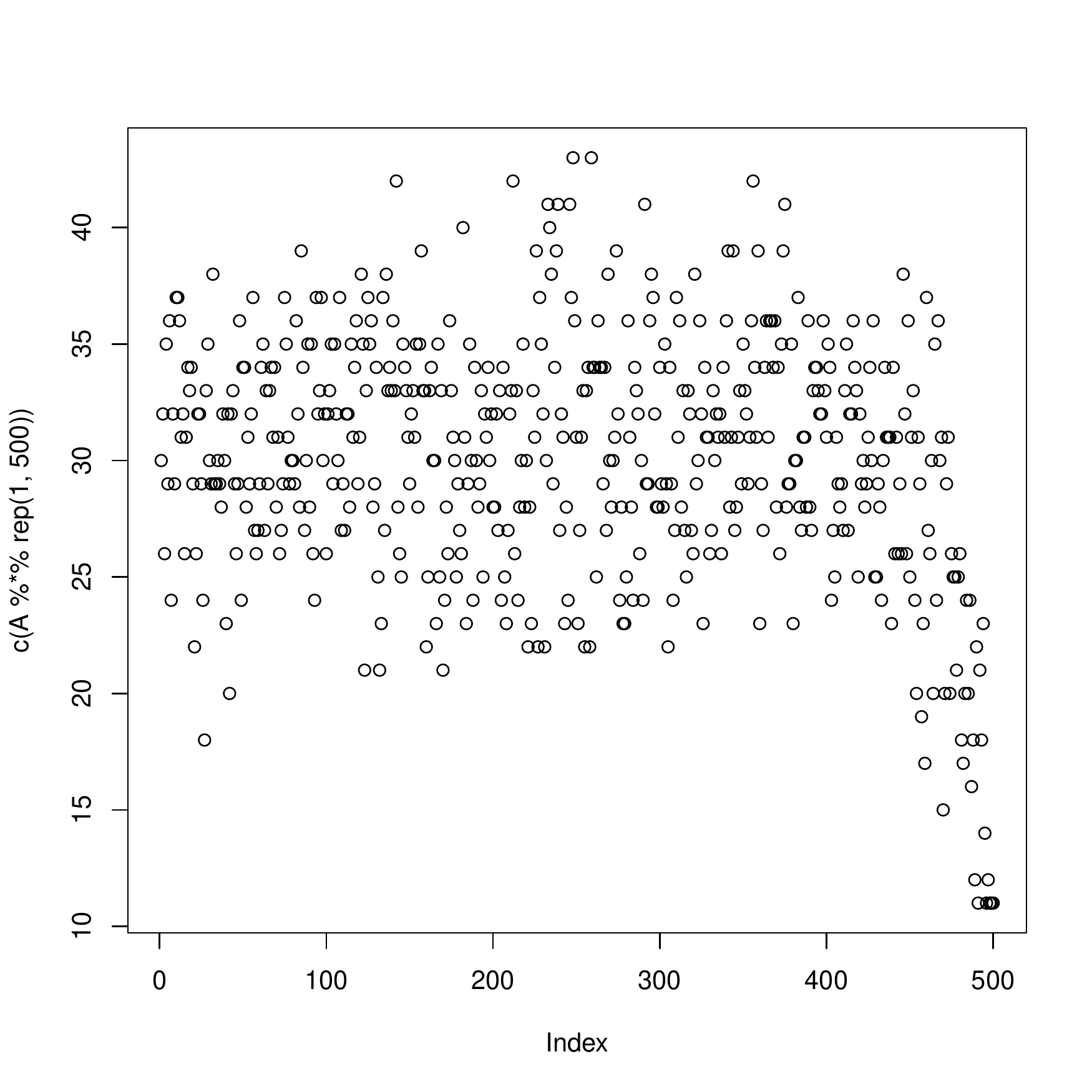}
  \caption{Coordinate ordering}
  \label{fig:connexion_coord_ordered}
\end{subfigure}
\caption{Number of connections of a point given its place in the ordering (same graphs as in figure \ref{fig:markov_connexions})}
\label{fig:markov_connexions_ordered}
\end{figure*}

The color count is overwhelmingly driven by the number of parents, the ordering, and interactions between them. 
The role of the parents is not surprising : the larger the parent sets, the more edges in the graph, the more colors needed. 
As for the ordering, it does not change the density but rather the distribution of the edges, which may explain why the number of colors is much smaller in the coordinate ordering.
In a graph obtained through Coordinate ordering and the Nearest Neighbor heuristic, a vertex tends to be connected with its immediate spatial surroundings. 
Indeed its parents in the DAG will be its predecessors along the coordinate used for ordering, its children will be its successors, and its co-parents will mostly be a mix of the closest parents and children.
In a graph obtained thanks to random or max-min ordering, the connections can be much longer, in particular for points coming early in the ordering. 
This results in some vertices being connected to many other vertices, leading to a denser graph. 
This point is illustrated in figures \ref{fig:markov_connexions} and \ref{fig:markov_connexions_ordered}.


The number of colors is robust with respect to the graph size because $n$ and its interactions have very low percentages in table \ref{table:sensitivity_coloring}. Since the numbers of colors are small with respect to $n$ ($45$ colors at the most), this makes chromatic sampling a good candidate for large data sets. 

This point is counter-intuitive because a bigger graph is more complex than a smaller graph and should therefore be more difficult to color. The markovian nature of NNGPs may be a lead to explain this point, because several sets of vertices that are not linked by a direct edge can be colored independently. While new vertices are added in the graph, older vertices can be forgotten and their colors can be re-used.

The dimension of the points $d$ plays almost no role in the sensitivity analysis, and choice of the coloring algorithm has a very marginal effect on the color count.

\begin{figure*}
\begin{subfigure}{.32\textwidth}
  \centering
  \includegraphics[width=1\linewidth]{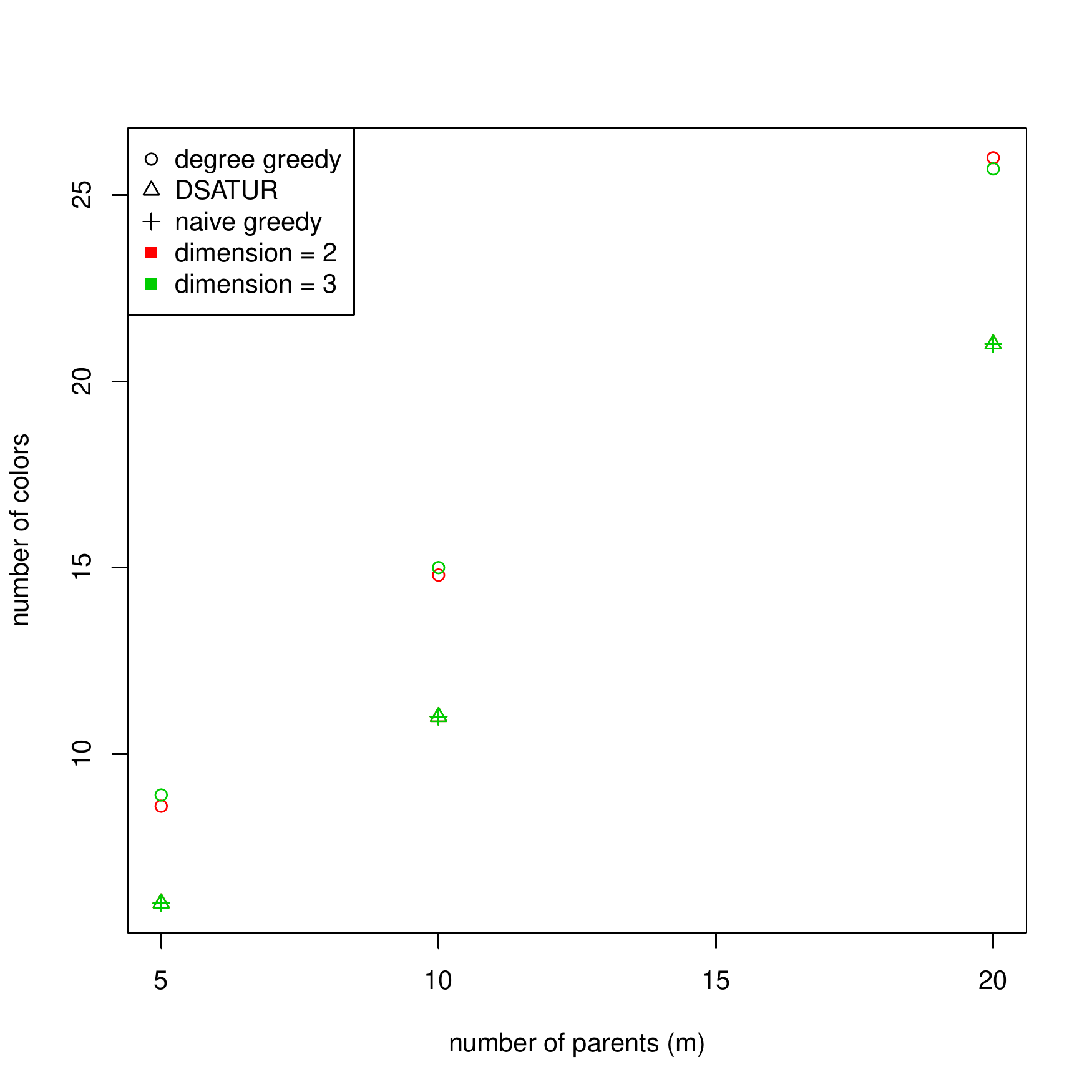}
  \caption{Coordinate ordering}
  \label{fig:algo_dim_coord}
\end{subfigure}
\begin{subfigure}{.32\textwidth}
  \centering
  \includegraphics[width=1\linewidth]{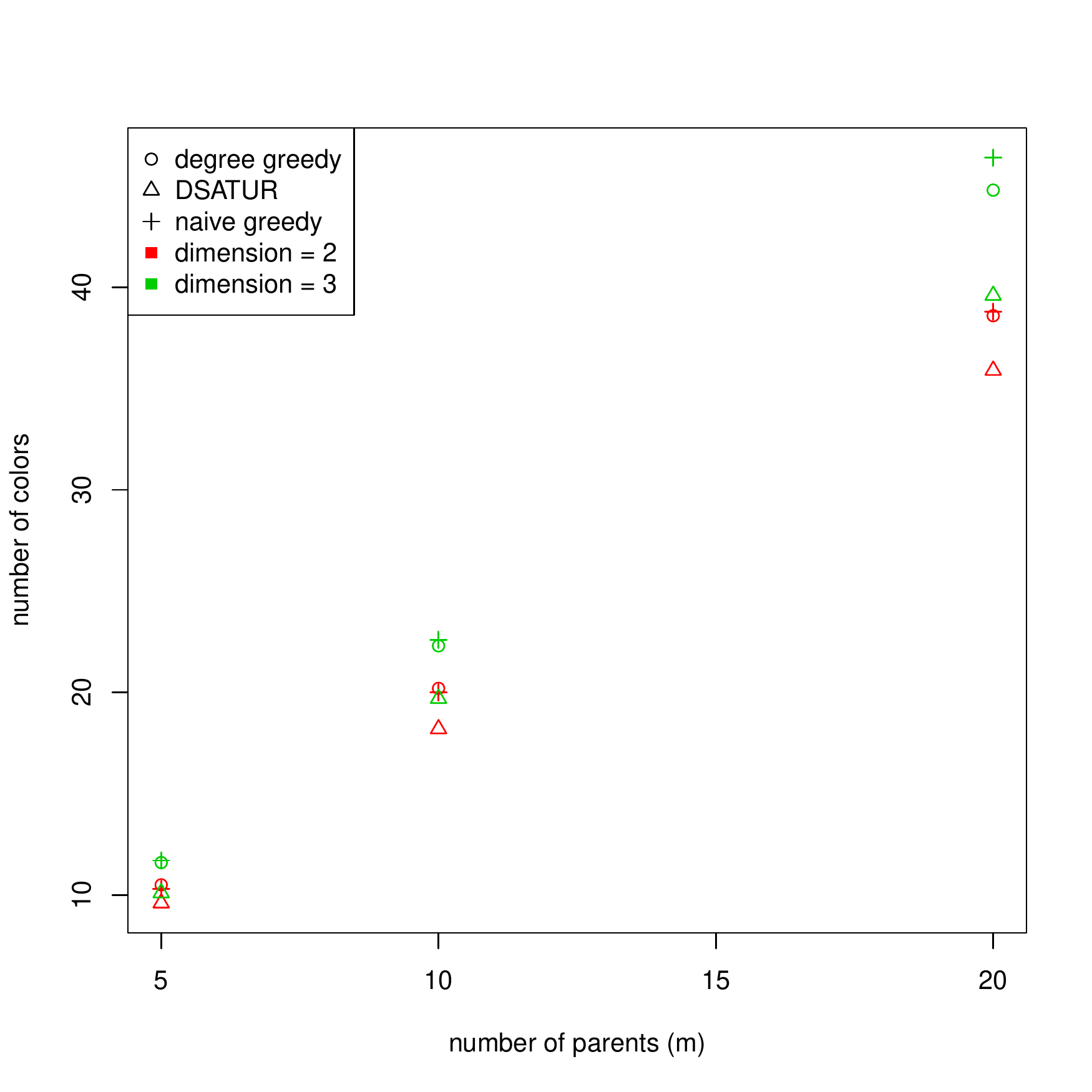}  
  \caption{Max-min ordering}
  \label{fig:algo_dim_maxmin}
\end{subfigure}
\begin{subfigure}{.32\textwidth}
  \centering
  \includegraphics[width=1\linewidth]{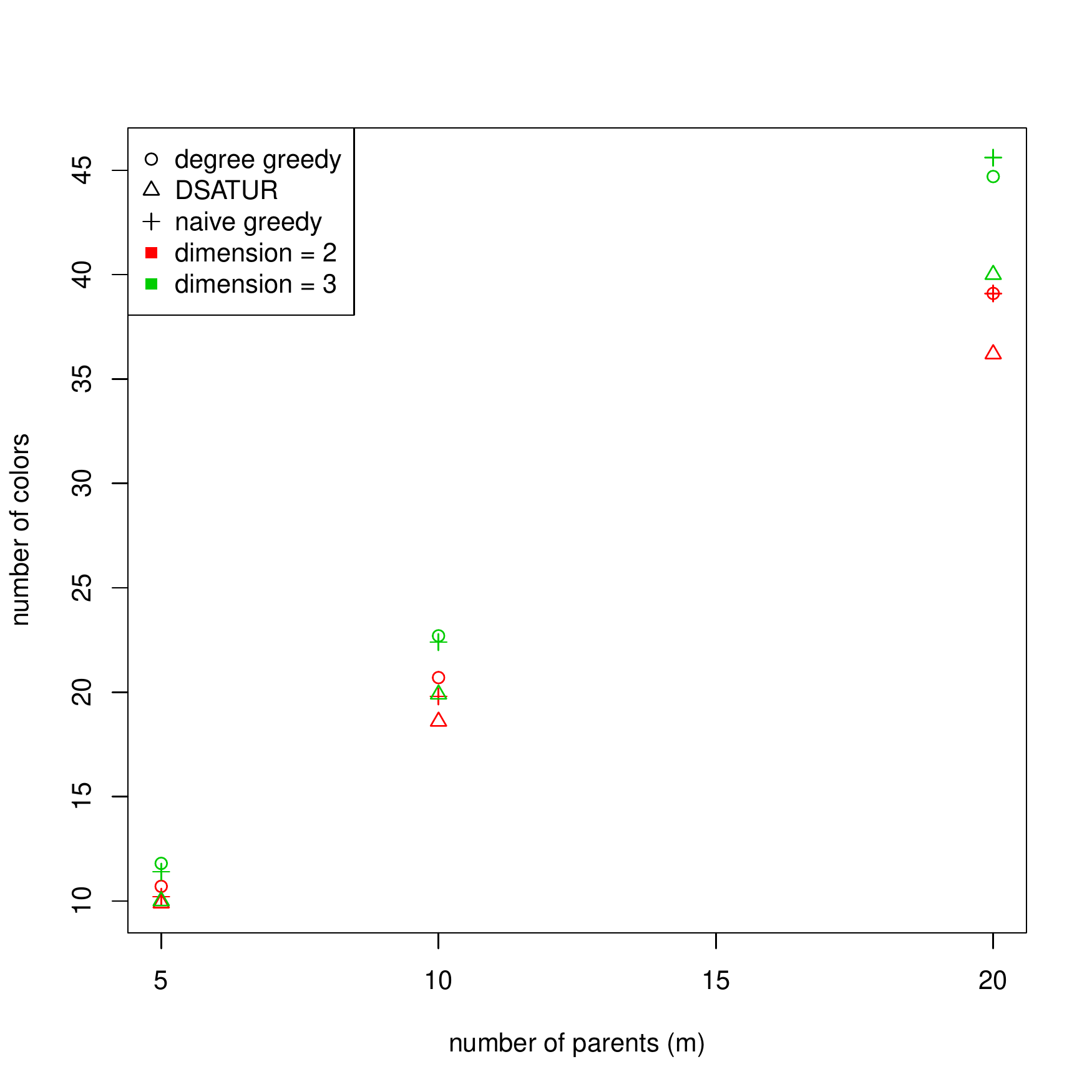}  
  \caption{Naive}
  \label{fig:algo_dim_random}
\end{subfigure}
\caption{Impact of the spatial domain dimension and the coloring algorithm on the mean number of colors, for graphs of size $n = 2000$.}
\label{fig:algo_dim}
\end{figure*}

Closer examination of the means reveals nonetheless that their effect is not nonexistent but rather dwarfed by the prominent role of the ordering of the vertices and the number of parents. 
For graphs obtained with max-min or random ordering (\ref{fig:algo_dim_maxmin} and \ref{fig:algo_dim_random}), the number of colors increases if $d=3$.

\begin{figure*}
\begin{subfigure}{.32\textwidth}
  \centering
  \includegraphics[width=1\linewidth]{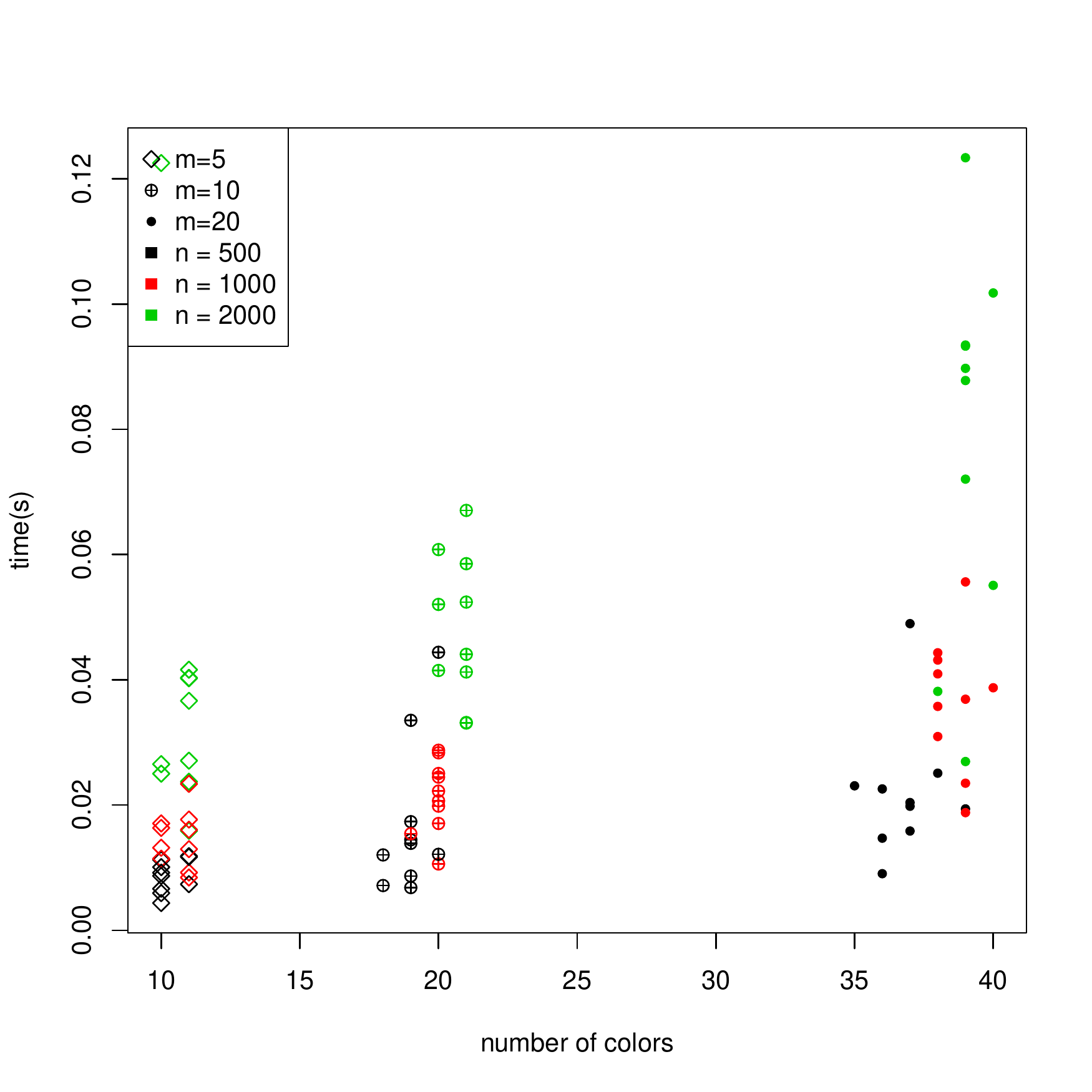}
  \caption{Degree}
  \label{fig:exp1_ncols_times_degree}
\end{subfigure}
\begin{subfigure}{.32\textwidth}
  \centering
  \includegraphics[width=1\linewidth]{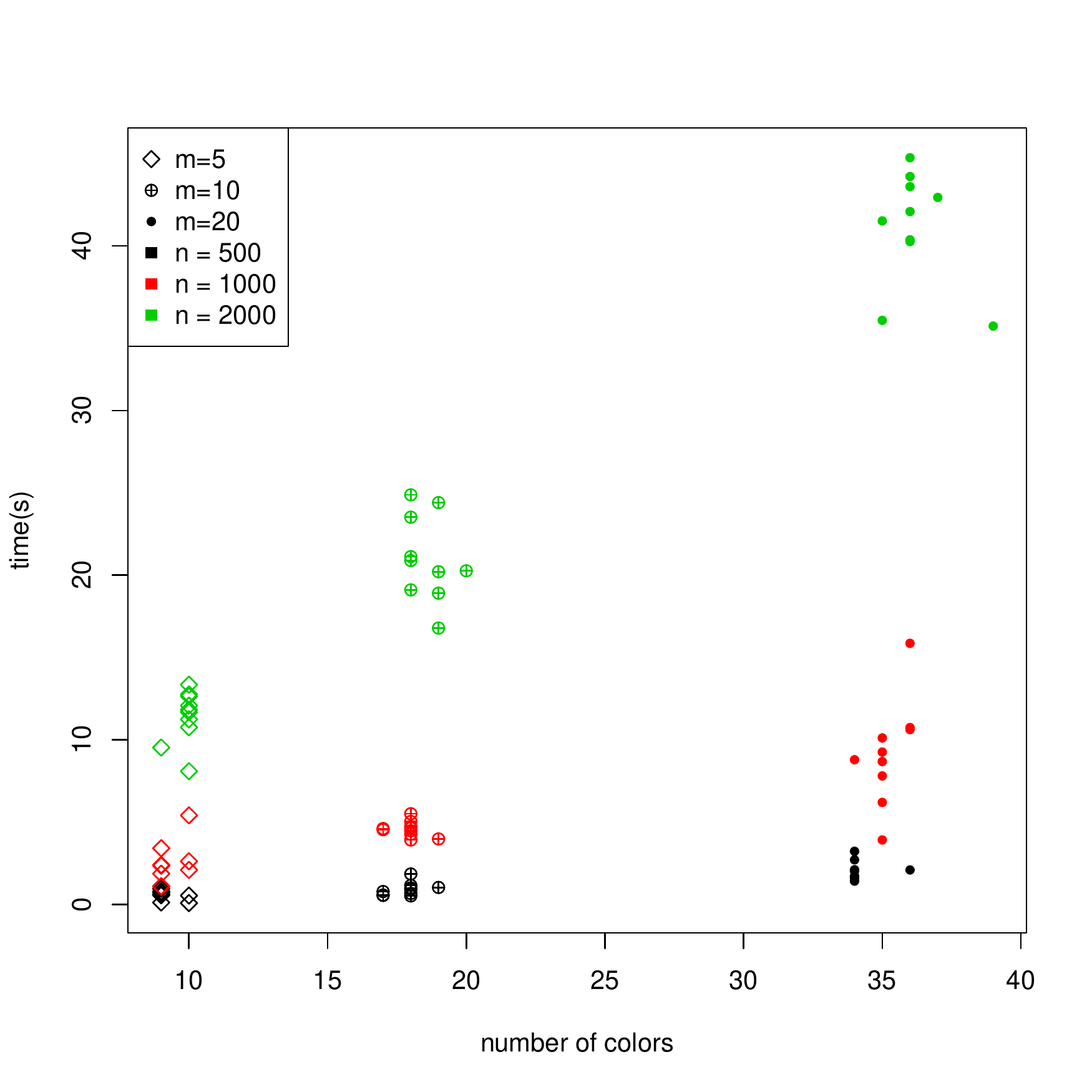}  
  \caption{DSATUR}
  \label{fig:exp1_ncols_times_DSATUR}
\end{subfigure}
\begin{subfigure}{.32\textwidth}
  \centering
  \includegraphics[width=1\linewidth]{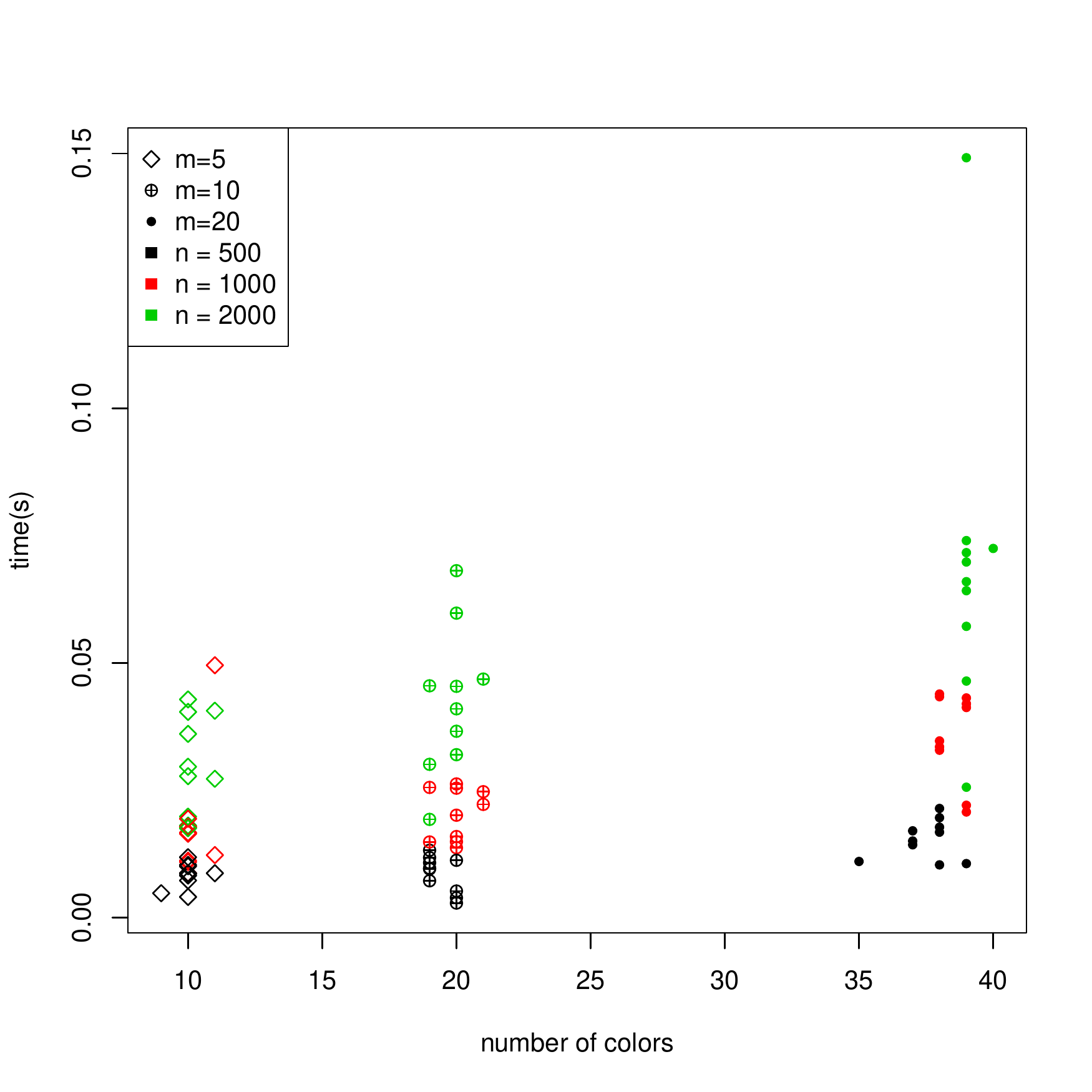}  
  \caption{Naive}
  \label{fig:exp1_ncols_times_naive}
\end{subfigure}
\caption{Repartition of the number of colors and running time with random ordering and d=2.}
\label{fig:exp_1_chromato}
\end{figure*}

The running time is affected by $n$, as expected. However, it is mostly explained by the coloring algorithm and its interactions with $n$ and $m$. 
In Figure \ref{fig:exp_1_chromato}, we see the results of the experiment when the ordering of the spatial points is random and $d = 2$. The number of parents $m$ defines well-separated vertical clouds of points, showing a clear, positive impact on the number of colors. It also increases the running time : the clouds of points on the right are stretched higher along the ordinates axis. The graph size $n$ affects the running time positively.
The other cases with different ordering and dimension all show this clear, chromatography-like profile. 

\begin{table}[ht]
\caption{Sensitivity analysis*}
\label{table:sensitivity_coloring}
\centering
\vspace{8pt}
\begin{tabular}{rrrrrrr}
& \multicolumn{2}{c}{Pilot}& \multicolumn{2}{c}{Large}& \multicolumn{2}{c}{Blocked}\\
              & colors& time& colors& time & colors & time \\
ordering      & 15.4 &  3.4 & 10.6  & 8.7   &  40.7 &  1.7 \\   
algo          &  0.8 & 24.7 & 0.5   & 1.3   &   1.0 &  8.6 \\  
d             &  0.6 &  0.0 & 1.1   & 0.7   &  0.4 &  0.0  \\  
m             &76.4 &  2.8 & 80.6  & 26.0  & 17.2 &  1.0  \\  
n/n blocks    & 0.2 & 11.8 & 0.1   & 40.7  & 15.2 & 16.2  \\  
ordering:algo & 0.3 &  6.8 & 0.1   & 0.3   &  0.3 &  3.3  \\  
ordering:d    & 0.3 &  0.0 & 0.5   & 0.4   &  0.2 &  0.0  \\  
ordering:m    & 5.3 &  0.8 & 5.2   & 5.4   &  7.9 &  0.5  \\  
ordering:n    & 0.0 &  3.3 & 0.0   & 3.0   &  6.4 &  6.1  \\  
algo:d        & 0.0 &  0.1 & 0.0   & 0.0   &  0.0 &  0.1  \\  
algo:m        & 0.2 &  5.5 & 0.2   & 0.6   &  0.1 &  1.9  \\  
algo:n        & 0.0 & 23.3 & 0.0   & 0.7   &  0.9 & 31.3  \\   
d:m           & 0.2 &  0.0 & 0.4   & 0.5   &  0.0 &  0.0  \\  
d:n           & 0.0 &  0.0 & 0.0   & 0.2   &  0.1 &  0.1  \\  
m:n           & 0.0 &  2.4 & 0.0   & 8.0   &  6.3 &  3.5  \\  
total         &99.7 & 84.8 & 99.5  & 96.6  & 96.6 & 74.6  \\   
\end{tabular}
\vspace{10pt}\\
\vspace{5pt}

\footnotesize{* Read : ``In the pilot experiment, the ordering of the spatial points explained 15.4 percents of the variance of the number of colors"}
\end{table} 

\underline{Benchmark}\\
In order to see if one coloring algorithm has the better of the others, we compare the average number of colors for each case of the experiment in table \ref{tab:table_means_exp1}. 
Regardless of the ordering, $m$, and $n$, the number of colors favors systematically but slightly DSATUR over the two simpler algorithms. In the case of coordinate ordering, naive greedy coloring reaches the performances of DSATUR. \\
While the two simple methods are very economical, the running time becomes high  in DSATUR when the graph size augments (Figure \ref{fig:exp1_ncols_times_DSATUR}).\\
We conclude that regardless of the structure of the graph, DSATUR must be chosen for smaller graphs. The two other methods must be chosen for larger graphs because DSATUR will become prohibitively expensive. 

\subsubsection{Coloring for large  graphs}
\underline{Design}\\
The objective is to test the sensitivity of the two interest variables and to benchmark coloring algorithms when the graphs are bigger. The experiment is the same as before, with two differences : 
\begin{itemize}
    \item Only naive greedy and degree greedy coloring algorithms are tested
    \item The graph size $n = 50000, 100000, 200000$
\end{itemize}
Each case is replicated $10$ times.

\underline{Sensitivity}\\
For the number of colors, the results are the same as before (table \ref{table:sensitivity_coloring}). It is mostly  determined by the ordering and the number of parents. The robustness of the number of colors with respect to $n$ is confirmed.
The running time is affected mostly by $n$, but the ordering and $m$ also play a role.

\underline{Benchmark}\\
In table \ref{tab:table_means_exp2}, we can see that naive coloring systematically has a lower mean number of colors than degree coloring. It is also slightly faster due to the fact that the vertices are not sorted. Anyway, the running times are short in both cases and are never bigger than 15 seconds. We conclude that naive greedy coloring is the better option for large data sets.

\subsubsection{Coloring blocked graphs}
\underline{Design}\\
The objective is to carry out sensitivity analysis and benchmark to graphs that correspond to spatial blocks used for block-update of the latent field. \\
Spatial clusters of vertices are found using a K-means algorithm on $n = 10000$ spatial locations, and coloring is applied to the Markov graph between the blocks. 
The orderings, the numbers of parents, and the dimensions remain the same as in the previous experiments. The parameters that change are :
\begin{itemize}
    \item The graph size $n_{blocks} = 10, 20, 50, 100, 500$.
    \item All three algorithms (DSATUR, naive greedy, and degree greedy) are tested
\end{itemize}
Each case is replicated $10$ times.

\underline{Sensitivity}\\
The sensitivity of the number of colors (table \ref{table:sensitivity_coloring}) differs from the previous experiments. Even though $m$ still matters, it is the ordering that becomes the most important variable. 

\begin{figure*}
\begin{subfigure}{.32\textwidth}
  \centering
  \includegraphics[width=1\linewidth]{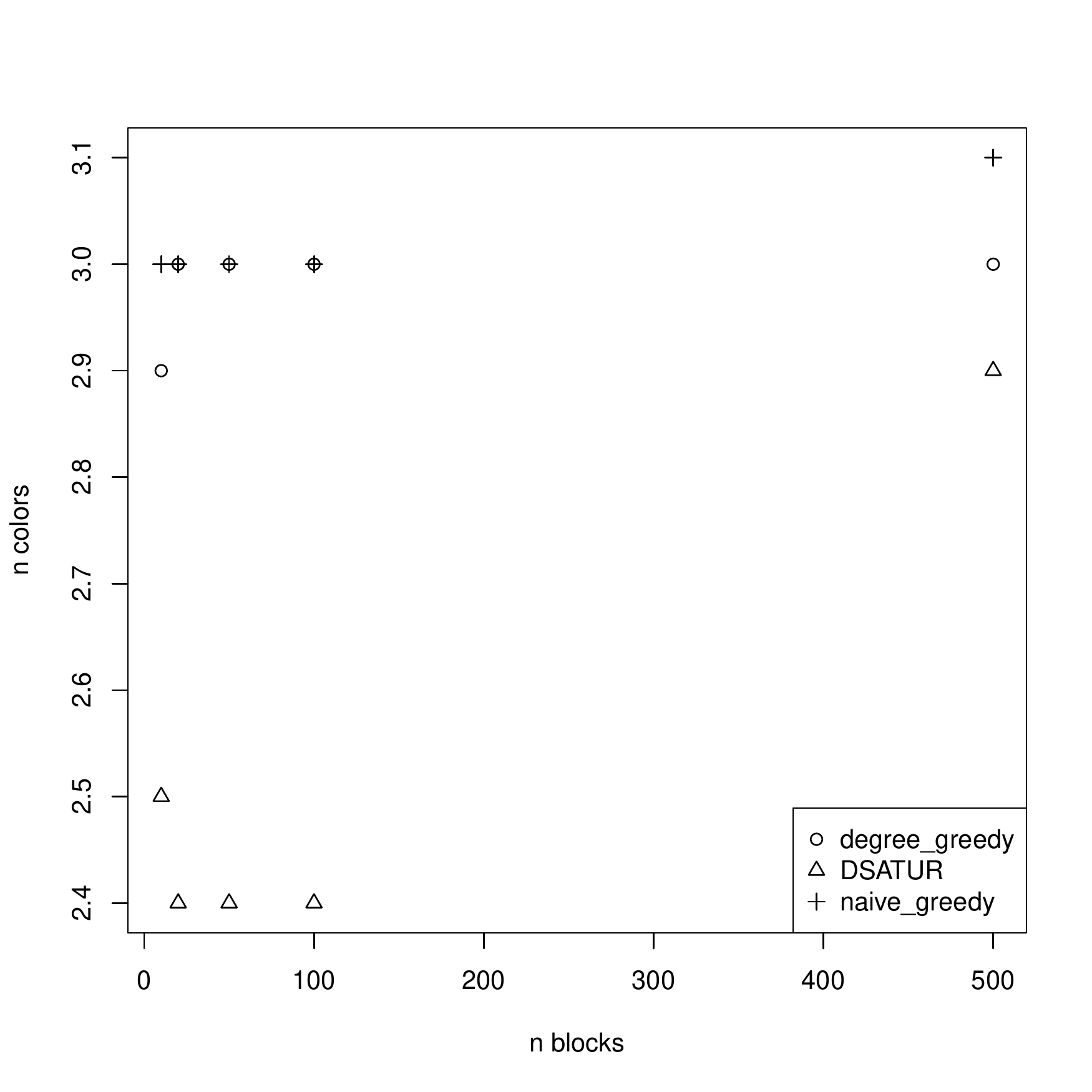}
  \caption{Coordinate ordering}
  \label{fig:n_blocks_n_colors_coord}
\end{subfigure}
\begin{subfigure}{.32\textwidth}
  \centering
  \includegraphics[width=1\linewidth]{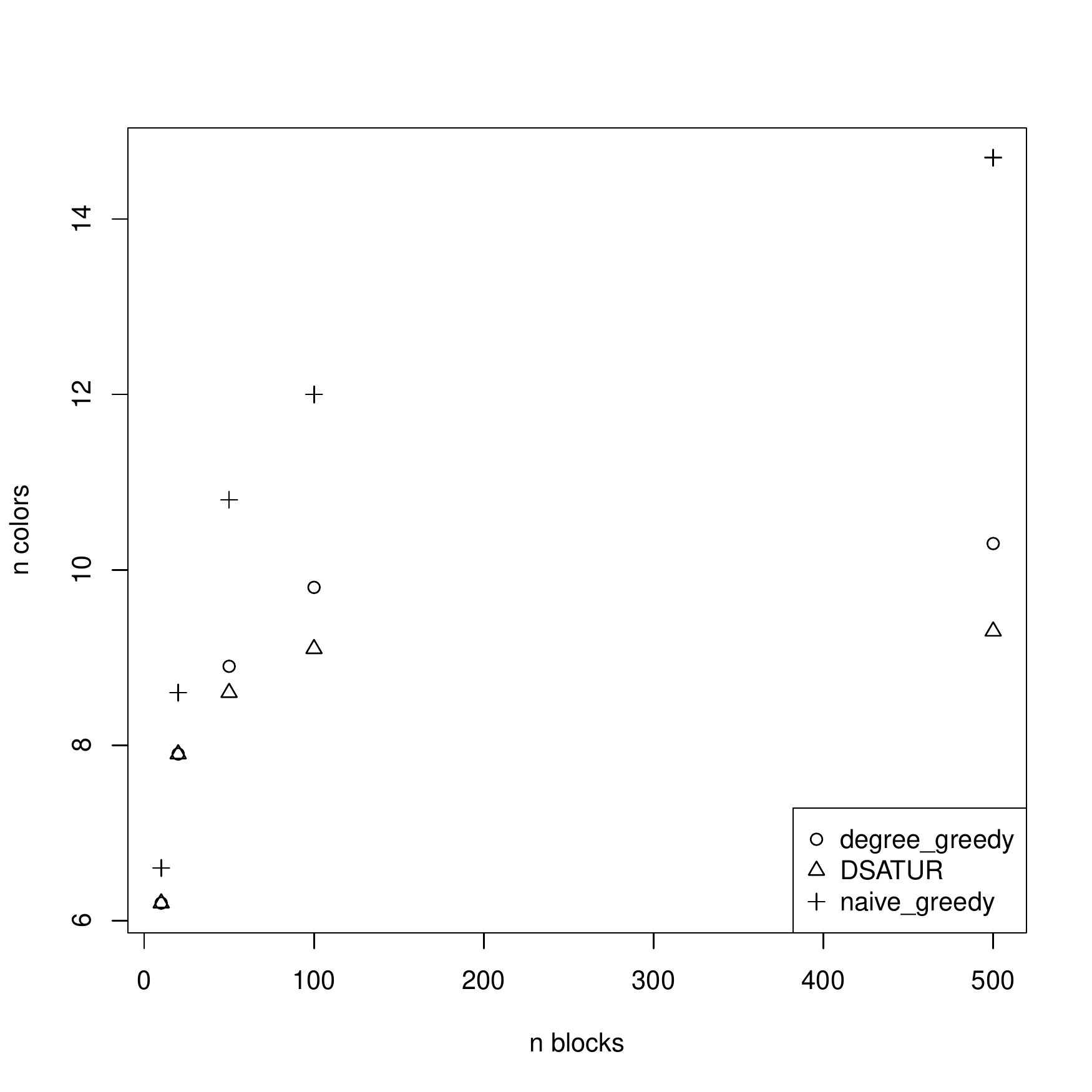}  
  \caption{Max-min ordering}
  \label{fig:n_blocks_n_colors_maxmin}
\end{subfigure}
\begin{subfigure}{.32\textwidth}
  \centering
  \includegraphics[width=1\linewidth]{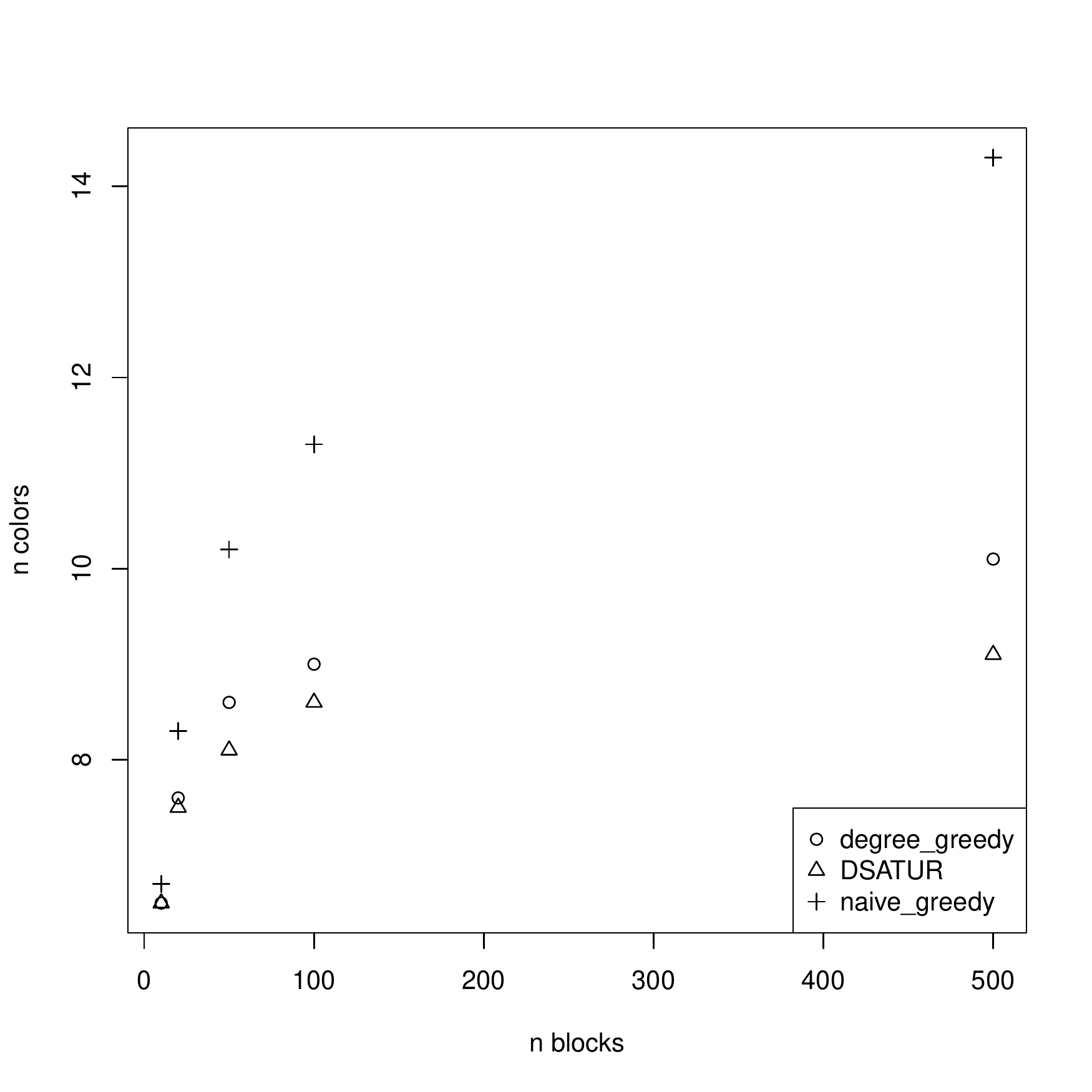}  
  \caption{Naive}
  \label{fig:n_blocks_n_colors_random}
\end{subfigure}
\caption{Number of colors following the number of blocks, for spatial domain dimension $d= 2$ and number of parents $m=5$.}
\label{fig:n_blocks_n_colors}
\end{figure*}

This loss of importance of $m$ can be explained by the fact that one edge is enough to connect two blocks, and once two blocks are connected adding new edges between them is redundant. 
On the other hand, the disposition of the edges in the space, which is induced by the ordering, keeps all its importance. Short edges induced by a coordinate ordering (\ref{fig:connexion_coord}) will connect  adjacent spatial blocks, while the long edges induced by the max-min heuristic (\ref{fig:connexion_maxmin}) will connect distant regions. 
The important interaction between $m$ ad the ordering is well explained by this hypothesis.
In table \ref{tab:table_means_exp3} we see that $m$ barely plays any role for coordinate ordering, while it keeps having an important impact for the other two orderings.
Indeed, when $\mathcal{S}$ is ordered following a coordinate, adding more short connections between contiguous spatial blocks does not change anything : those blocks already are connected. 
For max-min and random ordering, though, adding long edges may link distant regions that were not connected yet. 
After $m$ and the ordering, the number of blocks is the third most important variable. 
As expected, the more blocks in the graph, the more colors are needed. 
However, we remark that Max-Min and Random orderings perform poorly for graphs with few blocks, and actually need almost one color per block.
Once the graphs get bigger, the number of colors stabilizes. Therefore, the observed sensitivity with respect to the number of blocks is mostly induced by the bad coloring of graphs with few blocks.
The point can be visualized in figure \ref{fig:n_blocks_n_colors} for $m=5$ and $d = 2$. 

\underline{Benchmark}\\
Incontestably, DSATUR has the smallest number of colors, as seen in figure \ref{fig:n_blocks_n_colors}. Interestingly, degree greedy coloring has the second smallest number of colors.  If we assume that the number of blocks will always be smaller than 1000, we can discard the running time from our criteria and say that DSATUR is the best option for blocked graphs. However, in the cases with random and Max-Min orderings and low numbers of blocks, chromatic sampling does not greatly reduce the number of steps with respect to vanilla block sampling.

\section{Implementation, testing and application}
\label{sec:implementation}

\subsection{About our implementation}
\label{subsection:about_our_implementation}
We tested our implementation along with the state of the art package \textsf{spNNGP} presented in \cite{finley2017spnngp}, that uses the Gibbs sampler architecture given in \cite{NNGP}. \textsf{spNNGP} uses \textsf{Rcpp} \cite{eddelbuettel2011rcpp} and  parallelizes the computation of NNGP density.  In order to monitor convergence using the diagnostics from \cite{gelman1992inference, brooks1998general}, various chains need to be run one after the other. 
Our implementation is available at \url{https://github.com/SebastienCoube/Improving_NNGP_full_augmentation}. The code is done in  \textsf{R} (see \cite{R}), with the AS-IS Gibbs sampler architecture of \cite{yu2011center}. Chromatic sampling is implemented for individual locations. 
We used the package  \textsf{GpGp} (\cite{GpGp}) for Vecchia's approximation factor computation. Our implementation runs several chains in parallel thanks to the package \textsf{parallel}  (see \cite{parallel}), but \textsf{GpGp} does not implement parallel Vecchia's approximation factor computation within each chain like \textsf{spNNGP}. We emphasized the ease of use, with real-time Gelman-Rubin diagnostics and chains plotting, greedy MCMC tuning in the first hundred iterations, and the possibility to start, stop, and run again easily. 
For some data sets, our implementation has an advantage over \textsf{spNNGP} because multiple measurements at the same spatial site are allowed. However, unlike \textsf{spNNGP}, we have only implemented a Gaussian model so far.  

\subsection{Toy examples}
\label{subsection:toy_examples}
We present two toy examples in order to test our implementation, with the latent field NNGP implementation of \textsf{spNNGP}  as a reference. For both implementations, $5$ nearest neighbors were used for NNGP. 
The toy examples are Gaussian.
We compare the MCMC behavior using the number of iterations and the time needed before the chains have mixed following the Gelman-Rubin-Brooks $\hat R$.  We also compare the estimated covariance parameters with the values that were used to simulate the toy example. 
The covariance parameters are reported individually, and in the second toy example we report the Mean Square Error (MSE) of the fitted fixed effects with respect to their true value.
Eventually we compare the quality of the denoising using the MSE of the denoised field predicted by the model with respect to the simulated latent field. 
The first toy example is a simple Gaussian field simulated as follows. 
\begin{enumerate}
    \item Simulate spatial locations   $\mathcal{S}\sim \mathcal{U}([0, 50]\times [0, 50])$
    \item Simulate latent field  $w(\mathcal{S})\sim \mathcal{N}(0, \Sigma(\mathcal{S})), \Sigma(\mathcal{S})_{i, j} = exp(-0.5\|s_i, s_j\|)$ 
    \item Simulate observed variable  $z(\mathcal{S}) = w(\mathcal{S}) + \epsilon(\mathcal{S}), \epsilon(\mathcal{S})\sim \mathcal{N}(0, 5I_n)$ 
\end{enumerate}
\noindent The second toy example intends to highlight the positive effect of our architecture when covariates have some spatial coherence. We integrate covariates that are areal indicators, and others that are white noise.
\begin{enumerate}
    \item Simulate spatial locations   $\mathcal{S}\sim \mathcal{U}([0, 50]\times [0, 50])$ and note $\mathcal{S}_1$ the first coordinates of the locations
    \item Simulate latent field  $w(\mathcal{S})\sim \mathcal{N}(0, \Sigma(\mathcal{S})), \Sigma(\mathcal{S})_{i, j} = exp(-0.5\|s_i-s_j\|)$
    
    \item Simulate regressors $X = [X_1|X_2]$ with  $X_1 = [\mathbbm{1}_{1\leq \mathcal{S}_1< 2} | \mathbbm{1}_{2\leq \mathcal{S}_1< 3} \ldots |\mathbbm{1}_{49\leq \mathcal{S}_1\leq 50}]$ and $X_2$ a matrix of side $n\times 49$ with coefficients drawn following independent $\mathcal{N}(0, 1)$
    \item Simulate regression coefficients  $\beta \sim \mathcal{N}(0, I_{98})$
    
    \item Simulate observed variable  $z(\mathcal{S}) = w(\mathcal{S}) + X\beta^T+ \epsilon(\mathcal{S}), \epsilon(\mathcal{S})\sim \mathcal{N}(0, 5I_n)$ 
\end{enumerate}
The results of the runs on the toy examples are presented in table \ref{tab:chromatic_vs_spnngp}. 
The estimates are close to the target and there is no clear gap between the methods. 

\begin{figure}
\begin{subfigure}{.48\textwidth}
  \centering
  \includegraphics[width=.99\linewidth]{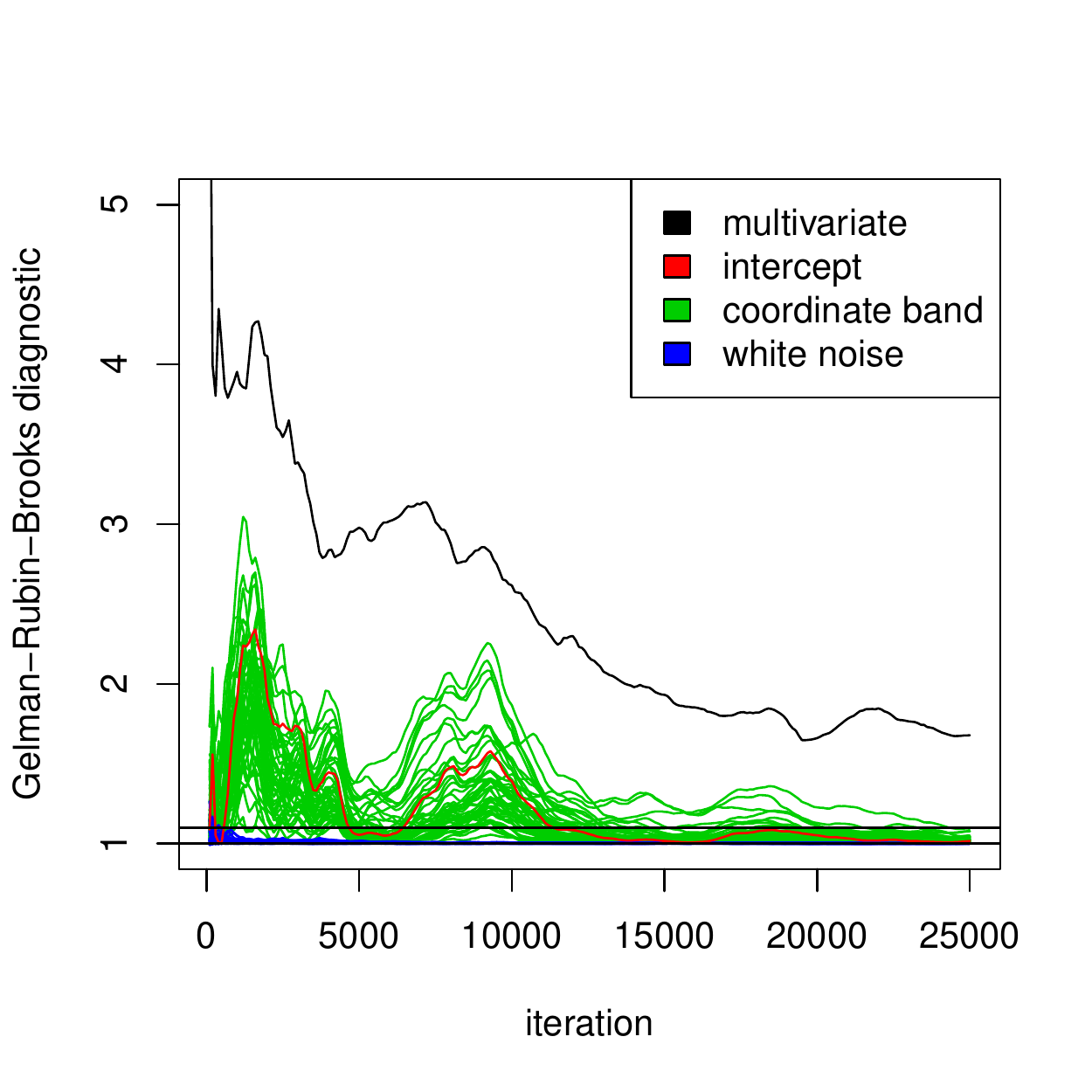}  
  \caption{\textsf{$\hat R$ with \textsf{spNNGP}}}
  \label{fig:GRB_beta_spNNGP}
\end{subfigure}
\begin{subfigure}{.48\textwidth}
  \centering
    \includegraphics[width=.99\linewidth]{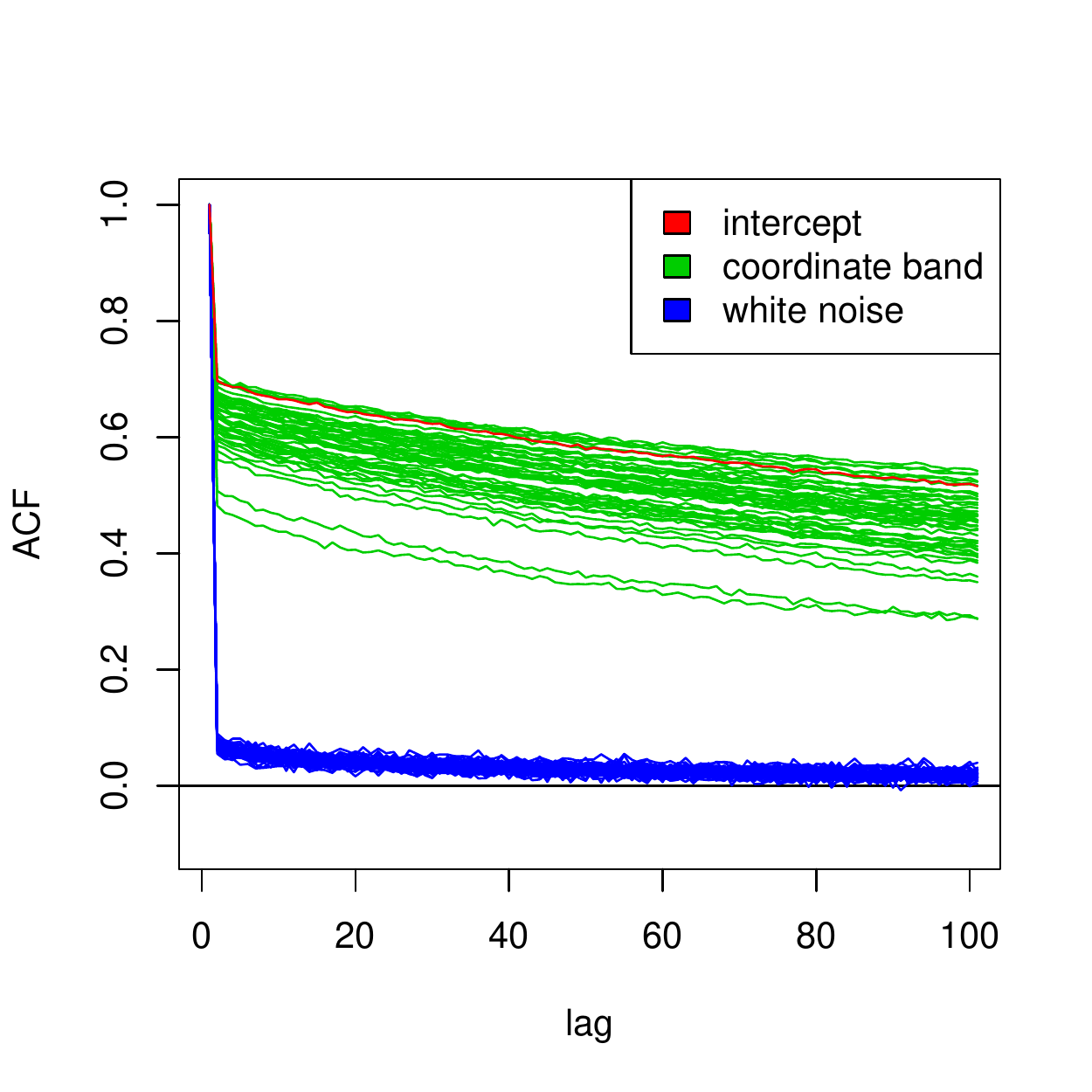}
  \caption{Autocorrelations with \textsf{spNNGP}}
  \label{fig:acf_beta_spNNGP}
\end{subfigure}

\begin{subfigure}{.48\textwidth}
  \centering
  \includegraphics[width=.99\linewidth]{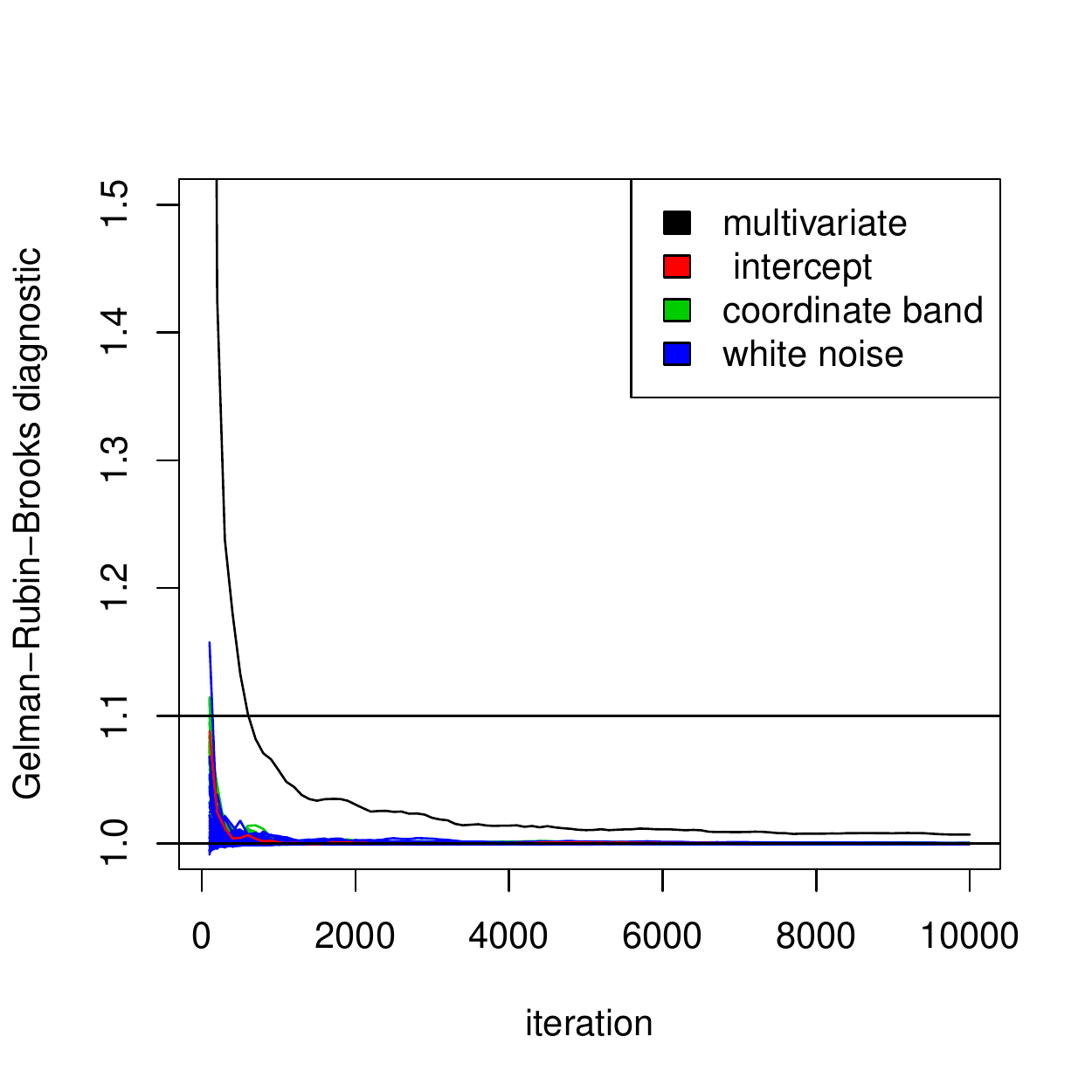}  
  \caption{\textsf{$\hat R$ with response \textsf{spNNGP}}}
  \label{fig:GRB_beta_spNNGP_res}
\end{subfigure}
\begin{subfigure}{.48\textwidth}
  \centering
    \includegraphics[width=.99\linewidth]{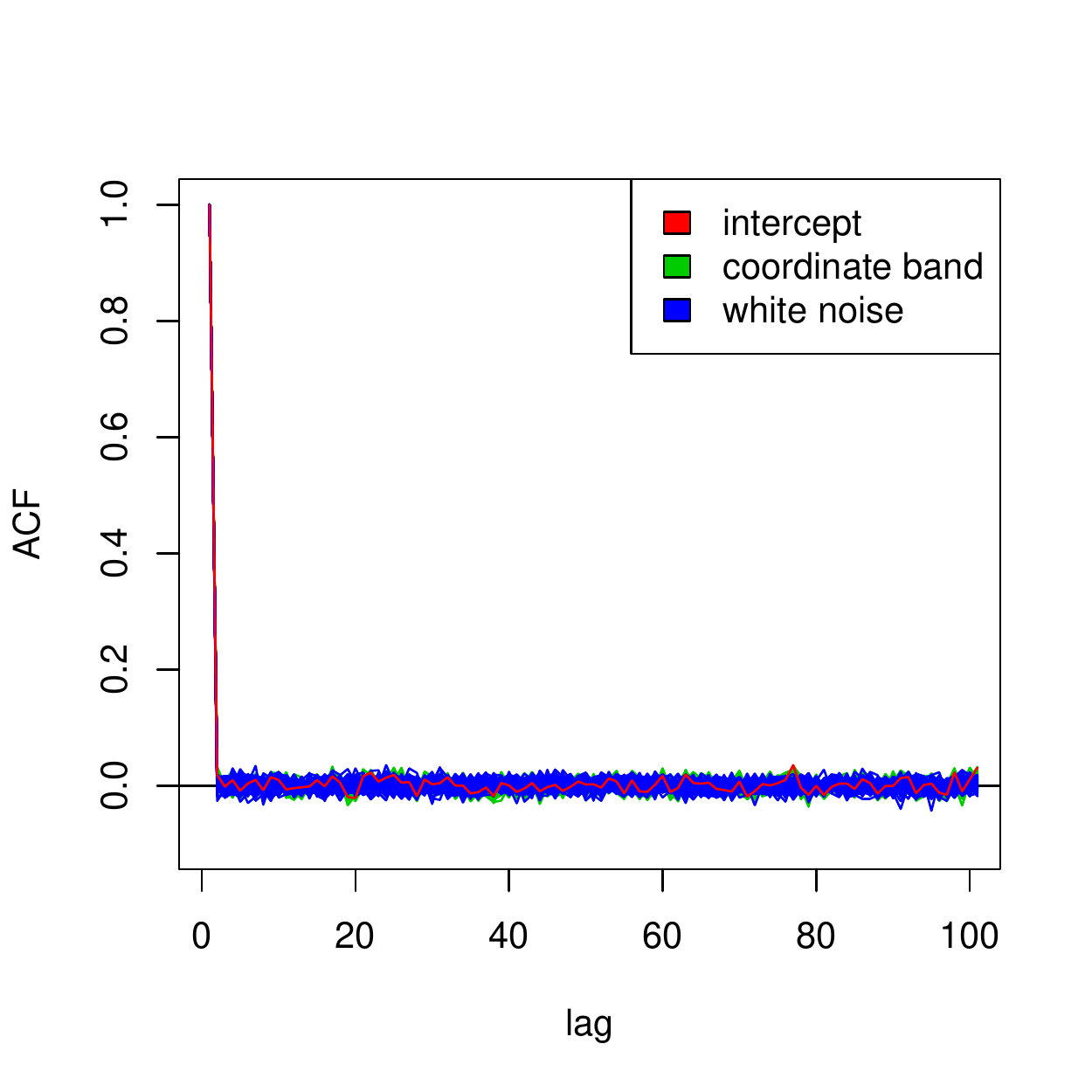}
  \caption{Autocorrelations with response  \textsf{spNNGP}}
  \label{fig:acf_beta_spNNGP_res}
\end{subfigure}

\begin{subfigure}{.48\textwidth}
  \centering
  \includegraphics[width=.99\linewidth]{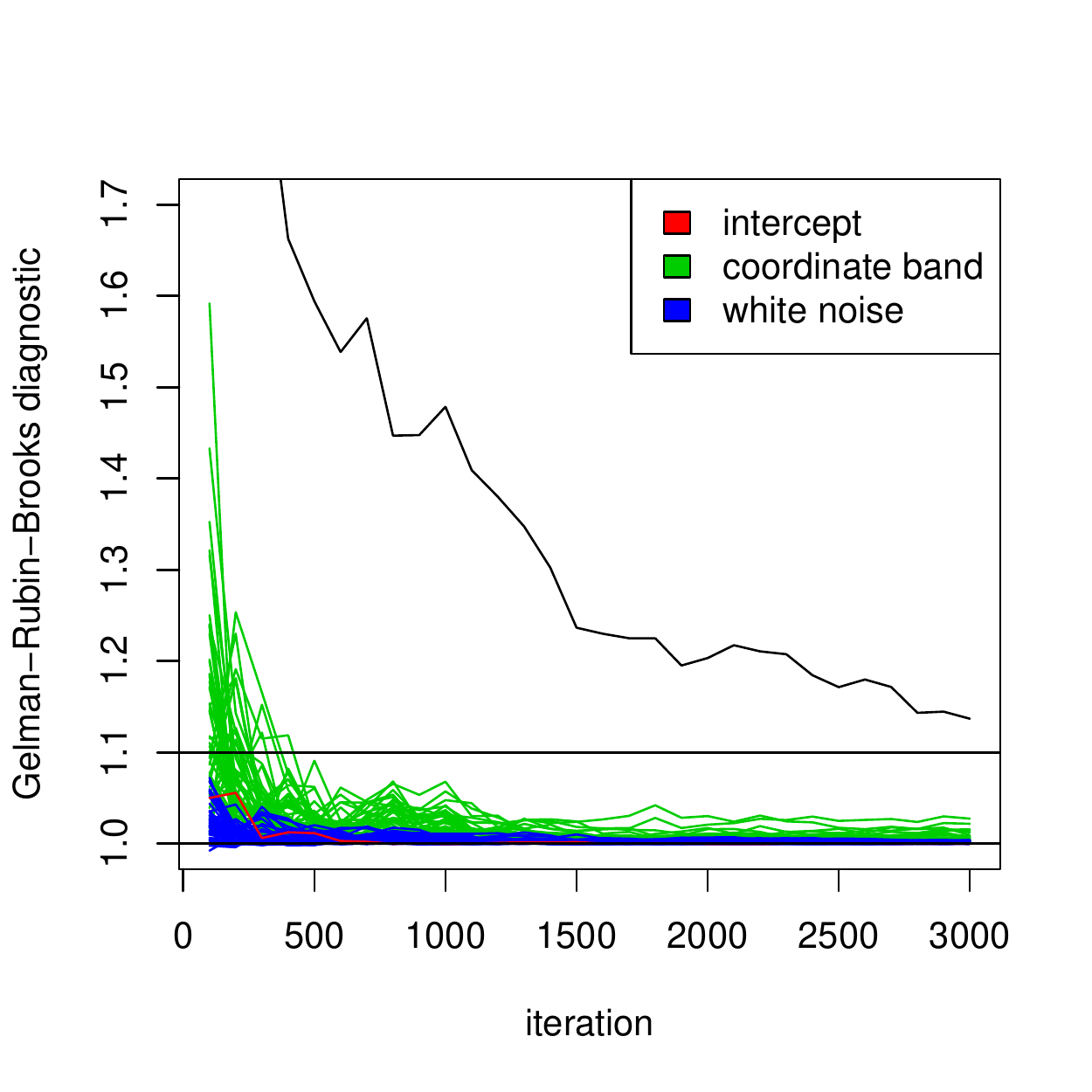}  
  \caption{\textsf{$\hat R$ of our implementation}}
  \label{fig:GRB_beta_interweaved}
\end{subfigure}
\begin{subfigure}{.48\textwidth}
  \centering
  \includegraphics[width=.99\linewidth]{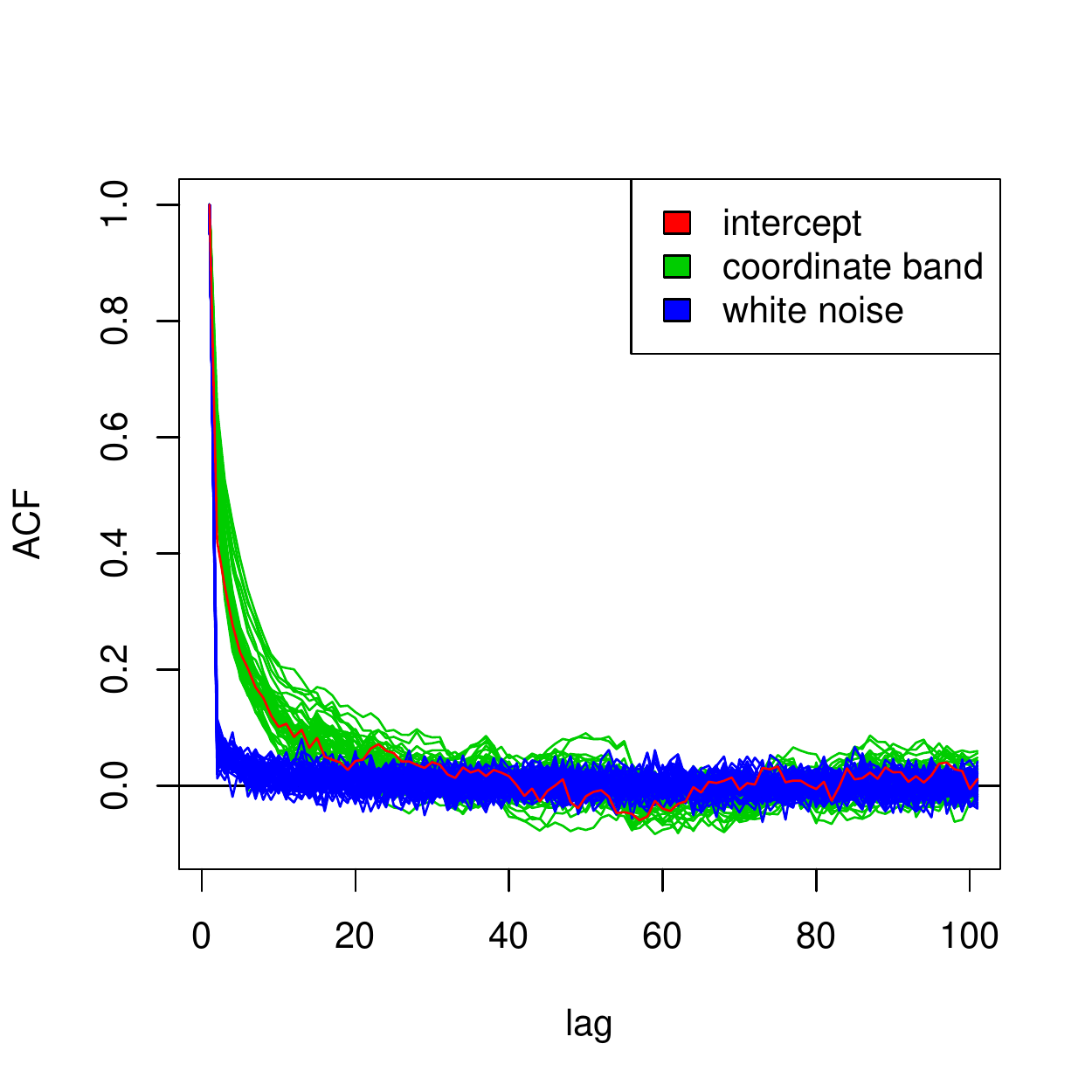}
  \caption{Autocorrelations of our implementation}
  \label{fig:acf_beta_interweaved}
\end{subfigure}

\end{figure}
\begin{figure}\ContinuedFloat
\begin{subfigure}{.48\textwidth}
  \centering
  \includegraphics[width=.99\linewidth]{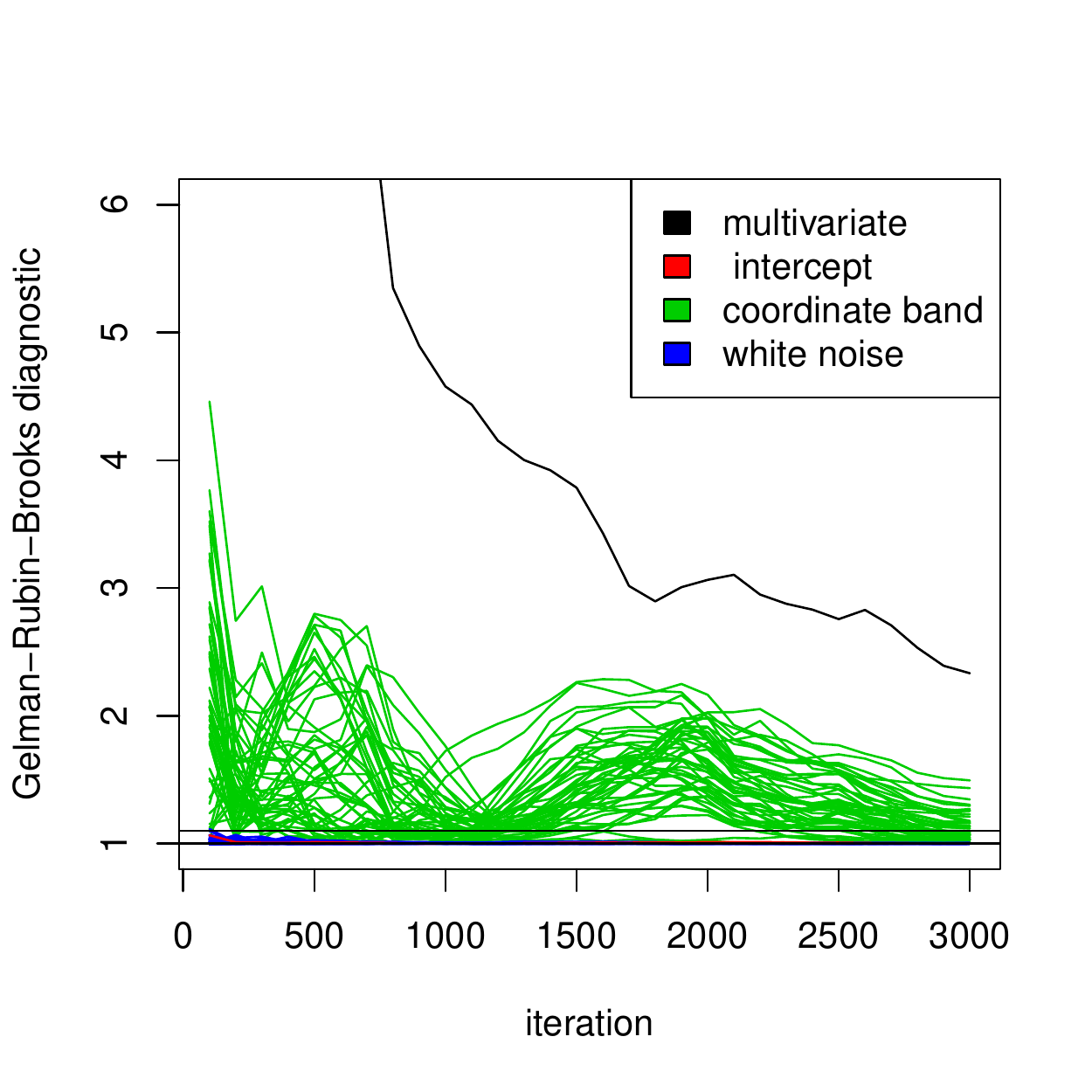}  
  \caption{\textsf{$\hat R$  of our implementation without interweaving}}
  \label{fig:GRB_beta_bad}
\end{subfigure}
\begin{subfigure}{.48\textwidth}
  \centering
  \includegraphics[width=.99\linewidth]{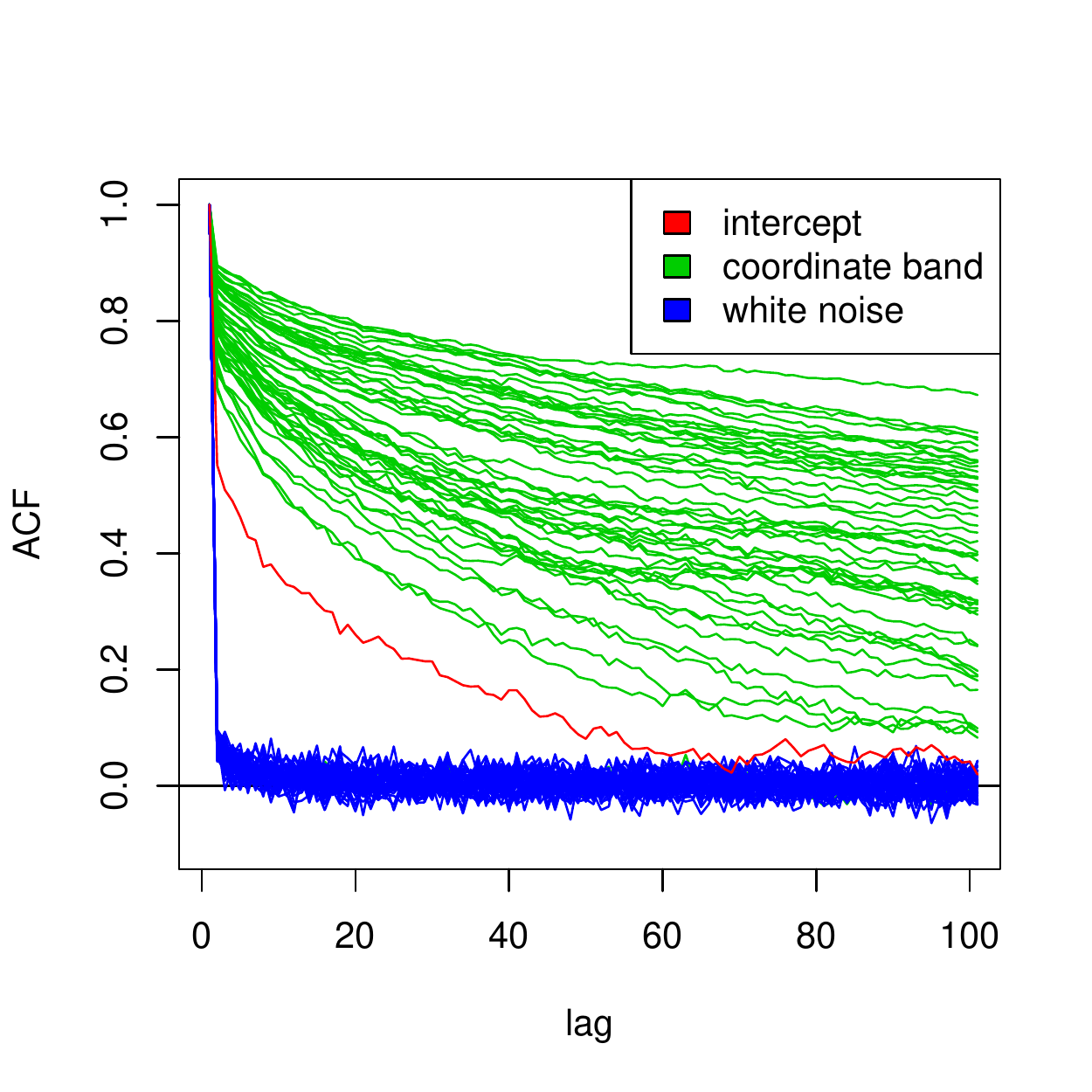}
  \caption{Autocorrelations  of our implementation without interweaving}
  \label{fig:acf_beta_bad}
\end{subfigure}

\caption{Behavior of the regression coefficients with spNNGP and with our implementation}
\label{fig:regression_coefficients_behavior}
\end{figure}

\begin{table*}
\caption{Summary of the toy examples runs}
\label{tab:chromatic_vs_spnngp}
\begin{subtable}{\textwidth}
\caption{Summary of the first toy example}
 \label{tab:chromatic_vs_spnngp_1}
\centering
\begin{tabular}{rrrrrrr}
  \hline
 method & n iter. & time (min) & MSE & latent var. & noise var. & range \\   \hline
spNNGP          &  15000 & 28 & 0.40 & 1.07 & 4.99 & 1.10 \\  
Our code        & 3000 & 28 & 0.38 & 1.08 & 5.00 & 1.01 \\ 
spNNGP res.   & 8000 & 13 &  & 1.06 & 5.00 & 0.99 \\ 
true values &    &   &   & 1.00  & 5.00 &1.00 \\ 
   \hline
\end{tabular}
\end{subtable}
\\

\begin{subtable}{\textwidth}
\caption{Summary of the second toy example}
\label{tab:chromatic_vs_spnngp_2}
\centering
\begin{tabular}{rrrrrrrr}
  \hline
method      & n iter. & time (min) & MSE  & $\beta-$MSE & latent var. & noise var. & range \\ 
spNNGP      & 25000   & 74 & 0.45 & 0.053 & 0.91 & 5.08 & 0.90 \\ 
Our code    & 3000    & 36 & 0.42 & 0.057 & 0.98 & 5.06 & 1.13 \\ 
spNNGP res. & 10000   & 50 &      & 0.047 & 0.88 & 5.10 & 0.72 \\ 
true values &         &           &     & & 1.00  & 5.00 &1.00 \\ 
   \hline
\end{tabular}
\end{subtable}
\end{table*}
Due to the fast-mixing  AS-IS architecture from \cite{yu2011center} and \cite{filippone2013comparative}, our implementation needed much less iterations than the latent model of \textsf{spNNGP} (Even taking into account the fact that one AS-IS iteration needs two covariance parameters updates) : our model takes thousands iterations to converge, while \textsf{spNNGP} needs tens of thousands.

The response model, in spite of its frugality,  needed a few thousands iterations to converge, like our implementation.  
The running times end up being of the same order, due to the efficient multi-process implementation of spNNGP which compensates the number of iterations. 
\\
Let's now focus on the behavior of the regression coefficients in the second toy example (Figure \ref{fig:regression_coefficients_behavior}). 
The best model regarding the mixing of the regression coefficients is incontestably the response model (\ref{fig:GRB_beta_spNNGP_res}, \ref{fig:acf_beta_spNNGP_res}). However, the covariance parameters needed more time to mix than the regression coefficients, explaining why $10000$ iterations were needed. 
Moreover, the response model cannot retrieve the latent field, explaining why its MSE could not be computed.
Except for the response model, we can see that the coherent regression coefficients of $X_1$, in green in \ref{fig:regression_coefficients_behavior}, mix slower than the fuzzy coefficients of $X_2$, in blue.
Nonetheless, for our implementation, the $\hat R$ diagnostics dropped to $1$ in a few hundred iterations (figure \ref{fig:GRB_beta_spNNGP}), against the tenths of thousands needed for \textsf{spNNGP} (figure \ref{fig:GRB_beta_interweaved}). 
For our implementation, the autocorrelations dropped to $0$ after a few dozen iterations  (figure \ref{fig:acf_beta_interweaved}).
The auto-correlations of \textsf{spNNGP} for the regression coefficients of $X_1$ were still between $0.4$ and $0.6$ after $100$ iterations (figure \ref{fig:acf_beta_spNNGP}), while the coefficients of $X_2$ remain stuck slightly above $0$. It is then clear that the chains behave much better in our implementation than in the \text{spNNGP}.
Moreover, the good behavior of our implementation could not be reproduced if we did not indicate that interweaving could be used, see figures \ref{fig:GRB_beta_bad}, \ref{fig:acf_beta_bad}.

\subsection{Application on to lead contamination analysis}
\label{subsection:lead_contamination}
We used our implementation to study a heavy metal  contamination data set proposed by Hengl in \cite{hengl2009practical}\footnote{https://spatial-analyst.net/book/NGS8HMC}. The dataset gathers measurements made by the United States Geological Survey of \cite{grossman2004national} and several covariates, including geophysical and environmental information about the sampling site, and potential contamination sources nearby. We added the predominant subsoil rock type given by the USGS study  presented in \cite{horton2017state}\footnote{https://mrdata.usgs.gov/geology/state/}. We scaled the quantitative regressors. 
After removing missing data, there was  $64274$ observations. 
We assumed the model 
$$log(z(s)) = w(s) + X(s) \beta^T + \epsilon (s),$$
$s$ being the sampling location, $X(\cdot)$ being the aforementioned covariates, $w(\cdot)$ being a latent Gaussian field with exponential covariance on the sphere, and $\epsilon$ being a white noise. 

The model converged in $4000$   iterations, and  1 hour and 38 minutes   were needed.  

We tried to analyze the real data set with \textsf{spNNGP} in order to compare the results and the running time. Surprisingly, \textsf{spNNGP} had a pathological behavior in spite of its good performances on simulated data. The scale parameter kept straying towards values several orders of magnitude above the variance of the observed field, even with starting points corresponding to our estimates. 
This behavior was observed with both latent and response model, and various orderings of the locations.

We present our implementation's estimates of the covariance parameters and some of the fixed effects in table \ref{tab:lead_contamination}. 
We left out some regressors such as the geological classification, indications about nearby mineral observations, the geophysical characteristics of the sampling site. 
The variances of the latent field and the noise have equivalent order ($\sigma^2 = 0.20 , \tau^2 = 0.18$). The spatial range is $30$ Km. With a rule of the thumb,  this means that the correlation drops to $10\%$  of the scale for locations separated by $60$ Km. The regressors behave as expected : the urbanization level and contamination indicators have a positive, certain effect on lead concentration. However, the values of the regression coefficients remain modest with respect to the scale of the latent field.

\begin{figure*}
\begin{subfigure}{\textwidth}
    \centering
    \includegraphics[width=\linewidth]{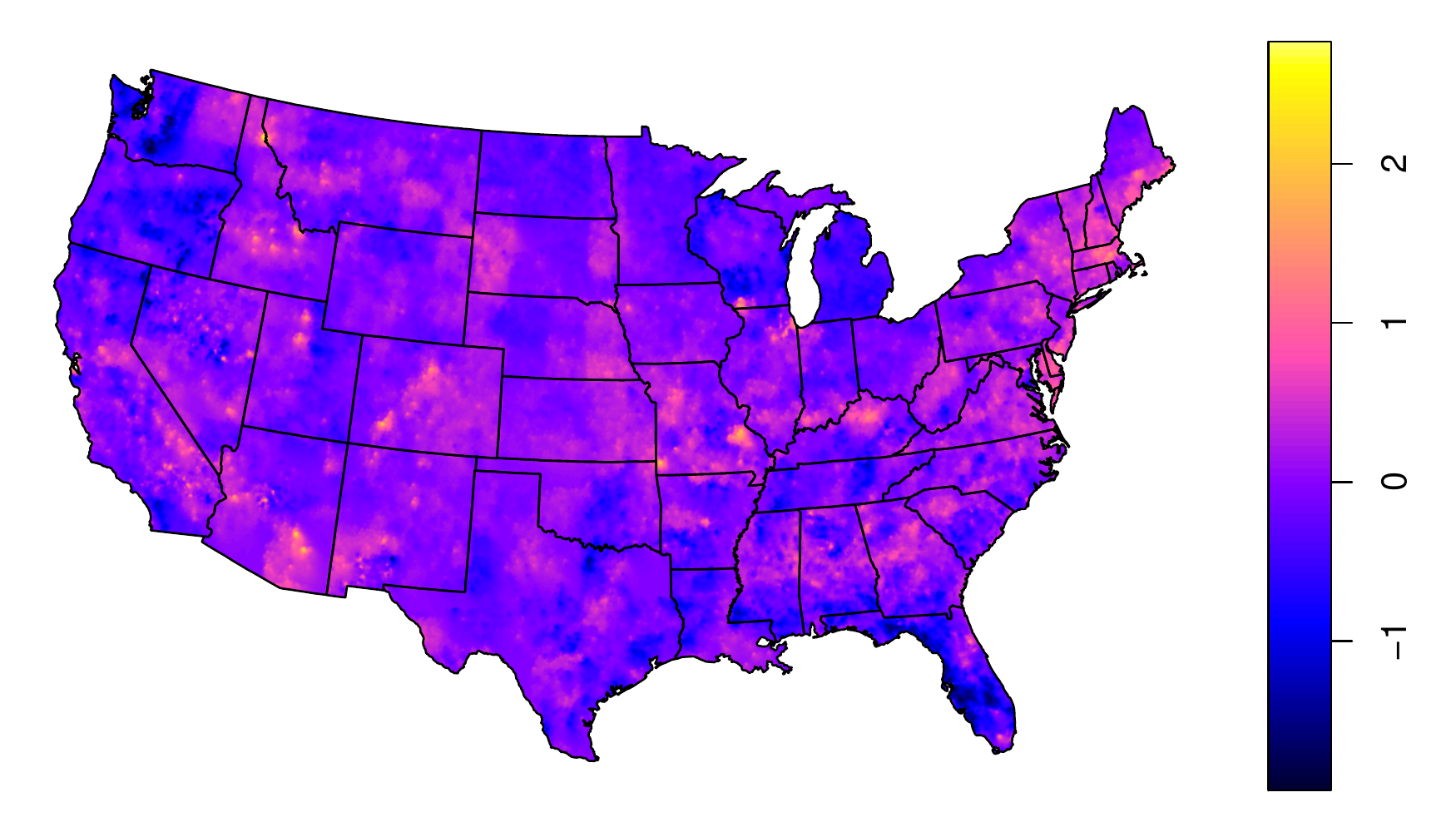}
    \caption{Predicted latent mean}
    \label{fig:lead_latent_field_mean}
\end{subfigure}
\begin{subfigure}{\textwidth}
    \centering
    \includegraphics[width = \linewidth]{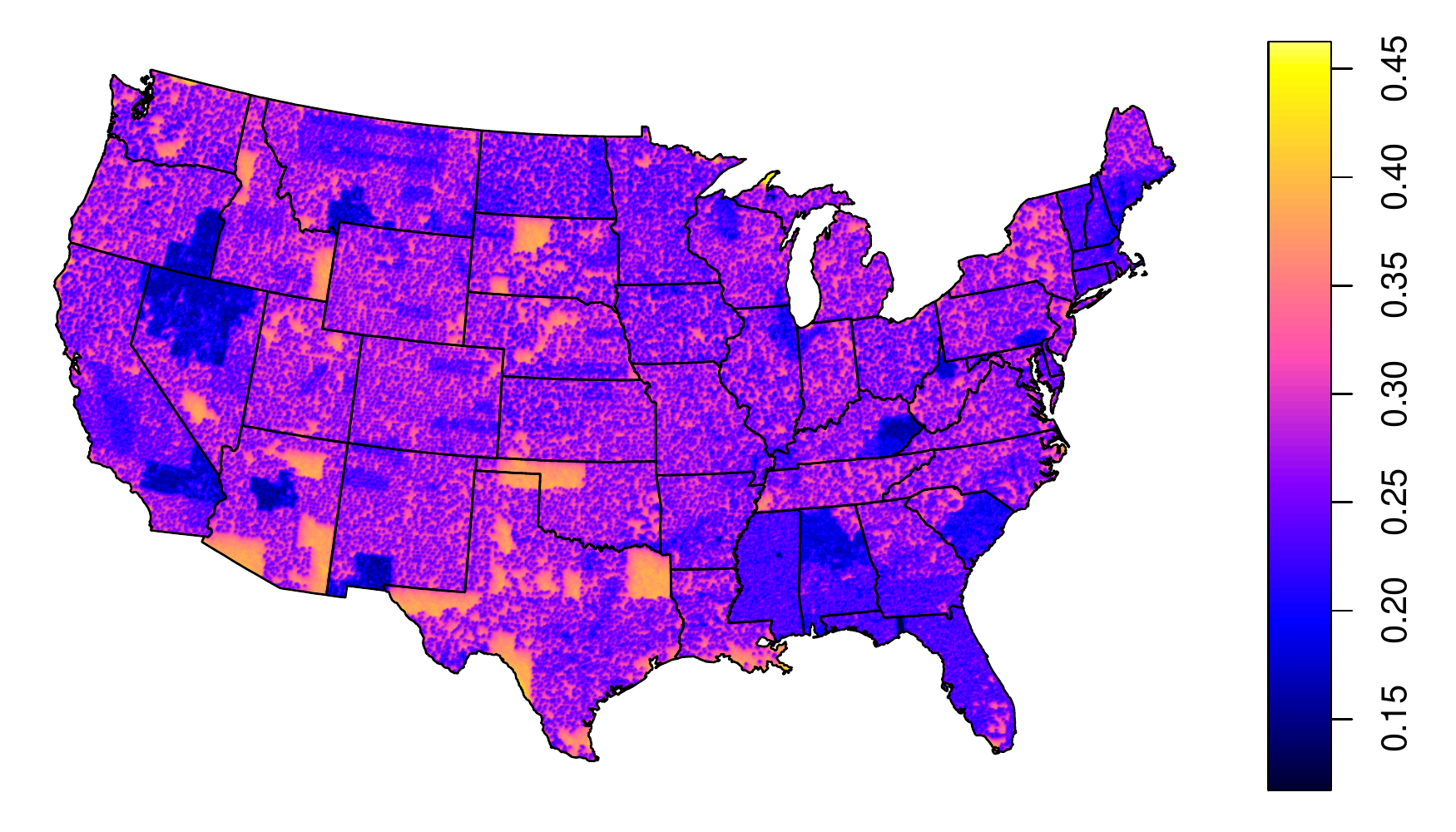}
    \caption{Predicted latent standard deviation (closely related to sampling density)}
    \label{fig:lead_latent_field_sd}
\end{subfigure}
\caption{Latent field predictions}
\end{figure*}

\begin{figure*}
    \centering
    \includegraphics[width=\textwidth,height=\textheight,keepaspectratio]{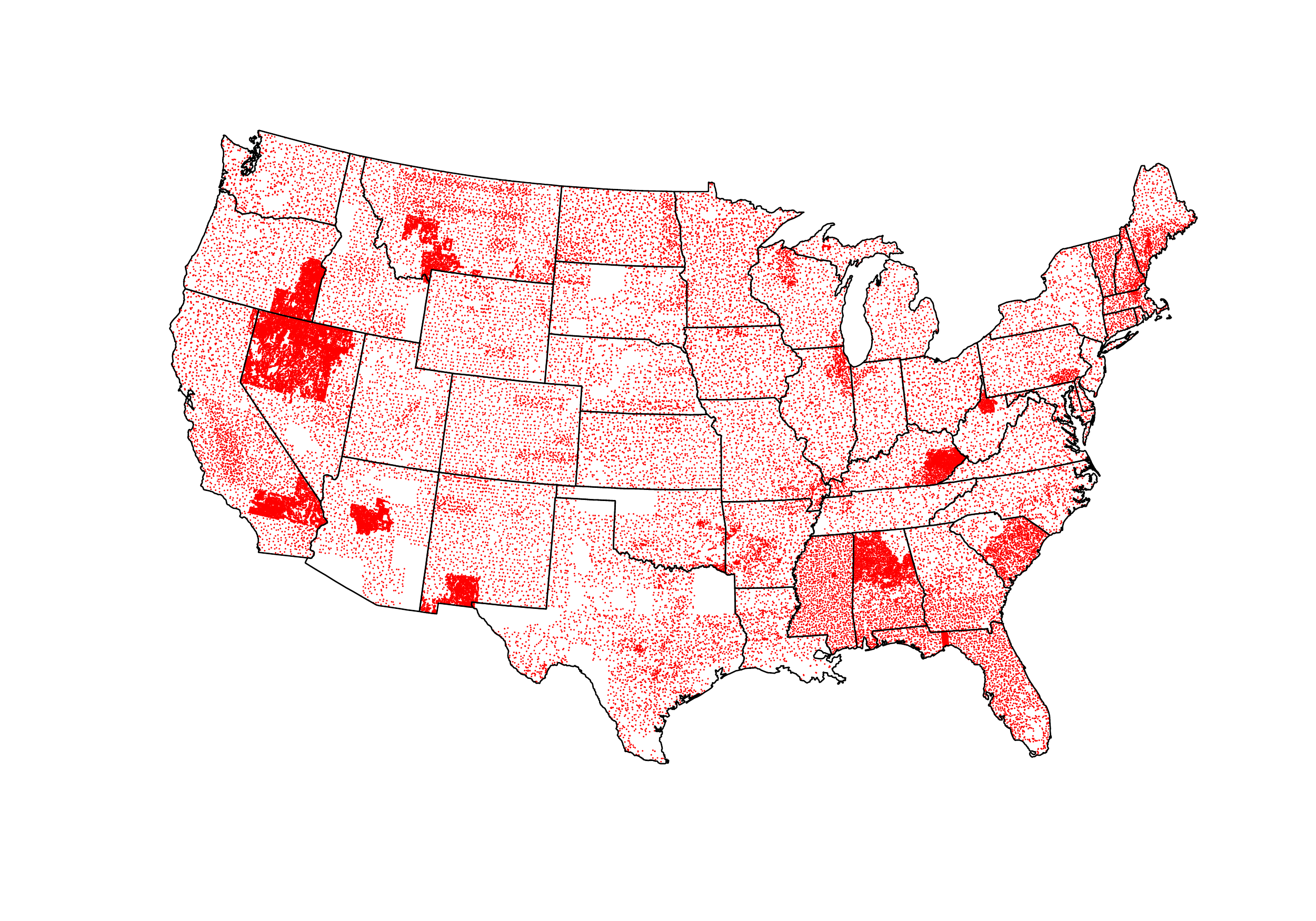}
    \caption{Sampling sites}
    \label{fig:sampling_sites}
\end{figure*}

We also provide prediction of the latent field on a $5$-Km grid on the territory of the USA mainland. Predictions  at un-observed locations are done  using the MCMC samples of the covariance parameters  $\theta$ and $w(\mathcal{S})$, see for example \cite{finley2019efficient}. 
We report the predicted latent mean and standard deviation in figures \ref{fig:lead_latent_field_mean} and \ref{fig:lead_latent_field_sd}. 
The standard deviation map must be put in relation with the sampling sites map (Figure \ref{fig:sampling_sites}).
The patches with high standard deviation correspond to zones with no measurement, while territories with dense sampling, such as Florida, will have low predicted standard deviation.

\begin{table*}[]
\caption{Summary of the covariance parameters and a subset of the fixed effects}
\label{tab:lead_contamination}
\centering
\begin{tabular}{rrrrrr}
  \hline
 & mean & qtile 0.025 & median & qtile 0.975 & st dev \\ 
 \hline
  Scale          & 0.198 & 0.188 & 0.198 & 0.209 & 0.0053 \\ 
  Noise variance & 0.178 & 0.175 & 0.178 & 0.181 & 0.0017 \\ 
  Range (Km)     & 35.6 & 33.3& 35.5 & 38.3 & 1.2700 \\

 \hline
  (Intercept) & 2.83 & 2.79 & 2.83 & 2.87 & 0.019 \\
  Air pollution dsty & 0.0403 & 0.0195 & 0.0404 & 0.0605 & 0.0106 \\
  Mineral operations dsty & 0.0180 & 0.0044 & 0.0182 & 0.0312 & 0.0069 \\ 
  Industrial toxic reject dsty & 0.0641 & 0.0433 & 0.0639 & 0.0852 & 0.0107 \\ 
  Carbon biomass dsty & -0.0543 & -0.0673 & -0.0541 & -0.0412 & 0.0066 \\ 
  Population dsty & 0.1360 & 0.1100 & 0.1360 & 0.1640 & 0.0138 \\ 
  Night light& 0.0384 & 0.0303 & 0.0384 & 0.0466 & 0.0042 \\ 
  Roads dsty & 0.0193 & 0.0139 & 0.0193 & 0.0248 & 0.0028 \\ 
   \hline
\end{tabular}
\end{table*}

\section{Discussion}
\label{sec:discussion}
We presented two ways to improve the behavior of NNGPs with full data augmentation, that can be simply applied to previous implementations.  What's more, while we assumed a Gaussian data model throughout the article, the two methods we proposed can be easily applied to other models.  \\
While our article focused on a basic NNGP model, our field centering may have applications in complex models. 
Space-varying regression coefficients are an extension to GP models (\cite{NNGP, PP}). 
If we consider the latent field $w(\cdot)$ as a space-varying intercept, it seems natural to try to center a spatially variable parameter on the corresponding fixed effect. The extension to other fixed effects we presented could prove valuable in the case in which the regressor with a spatially variable $\beta$ is correlated with other variables from $X$.  
Another possible extension could be a GP defined as the sum of two or more GPs. It could have an interest in various applications, such as : modelling seasonality in a space-time process, modelling a process with short-range and long-range interactions, defining one non-separable space-time process as a sum of two separable processes.  The equivalent of standard parametrization would be $z(\cdot) = \beta_0 + w_1(\cdot) + w_2 (\cdot) + \epsilon$, $w_1(\cdot)$ and $w_2(\cdot)$ being GPs of mean 0. One could try out a Russian doll centering :   $z(\cdot) = v_1(\cdot) + \epsilon$ where $v_1(\cdot)$ has mean $v_2(\cdot)$, and $v_2(\cdot)$ has mean $\beta_0$. In this case, in addition to apply the extension to other fixed effects we presented, it might be necessary to interweave the standard and the ``Russian doll'' parametrizations. 
\\
Beyond the improvements of chromatic sampling in the NNGP algorithm, exploration of the moralized graph could be an interesting approach to study Vecchia's approximation and evaluate heuristics concerning ordering and picking parents. 
For example, Guinness \cite{Guinness_permutation_grouping} has explored how various ordering and grouping strategies affected the Kullback-Leibler divergence of Vecchia's approximation with respect to the full GP density. 
Those strategies have a graphical translation. Grouping takes an existing graph and adds new edges, making it closer to the full GP's graph (i.e. the saturated DAG and moralized graph). Ordering modifies the structure of the graph and the length of the edges, just like the mixing of observations explored in  \cite{stein2004approximating}. For example, a coordinate or a middle-out ordering with Nearest Neighbor heuristic will make graph where each vertex connected to its closest neighbors, while we could use a classical concept of Geography and say that a random or a max-min ordering will generate graphs not unlike a Christallerian system. 
Focusing on the neighbor-picking heuristics gives one a close-up shot of what is going on and has a direct algorithmic translation, but some descriptive statistics about the moralized graphs could give a more general view. \\

\bibliographystyle{chicago} 
\bibliography{references.bib}

\clearpage
\appendix

\section{Stochastic form of the intercept-field model}
\label{section:stochastic_form_centering}
We obtain the full conditionals of $w_c$ and $w_s$ using the conditional expectation and variance formulas using precision matrices of \cite{GMRF_Rue_Held}. The joint precision of either $w_s$ or $w_c$ being 
$
\left[
\begin{array}{cc}
     \tilde Q+ \tau^2I_n&-\tau^2I_n  \\
     -\tau^2I_n&\tau^2I_n\\ 
\end{array}
\right] 
$. The expectation of $w_s$ is $0$, the expectation of $w_c$ is $\beta_0$, and the expectation of $z$ is always $\beta_0$.

\underline{Distributions with the standard model}\\
The full conditional distributions of $\beta_0$ and $w_s$ are : 
$$[\beta_0 | w_s]\sim \mathcal{N}(\overline{z-w_s}, \tau^2/n),  [w_s|\beta_0]\sim\mathcal{N}( -(\tilde Q + I_n/\tau^2) ^{-1}(-I_n/\tau^2)(z-\beta_0),(\tilde Q + I_n/\tau^2) ^{-1}).$$
Note $\textbf{1}$ the vector of length $n$ and filled with ones. 
 From the second full conditional, we have a formula for the mean of $w_s$, which is obtained with $\overline{w_s} =\textbf{1}^tw_s/n$. It has 3 terms : one is fix, the second is a geometric carry-over of $\beta_0$, and the third is stochastic : 
$$ [\overline{w_s}|\beta_0]\sim 
\underbrace{\textbf{1}^T (\tau^2\tilde Q + I_n) ^{-1}(z)/n}_{\text{fixed}}  -
\underbrace{(\textbf{1}^T (\tau^2\tilde Q + I_n) ^{-1}\textbf{1}/n) (\beta_0)}_{\text{carry-over}}  +
\underbrace{\mathcal{N}((0,\textbf{1}^T(\tilde Q + I_n/\tau^2)^{-1} \textbf{1}/n^2)}_{\text{innovation}}.
$$
Injecting the full conditional of $\bar{w_s}$  into $\beta_0$'s, we identify an expression with 3 terms like before : 
$$
\begin{array}{ccl}
[\beta_0^{t+1}|\beta_0^t] &\sim &
\underbrace{\overline{z} - \textbf{1}^T (\tau^2\tilde Q + I_n) ^{-1}(z)/n}_{\text{fixed}} + \underbrace{(\textbf{1}^T (\tau^2\tilde Q + I_n) ^{-1}\textbf{1}/n)  \beta_0^t}_{\text{carry-over}}\\
     && 
+  \underbrace{\mathcal{N}(0,\textbf{1}^T(\tilde Q + I_n/\tau^2)^{-1} \textbf{1}/n^2 + \tau^2/n)}_{\text{innovation}}. 
\end{array}
$$ 

\underline{Distributions with the centered model}\\
The full conditional of $\beta_0$ and $w_c$ are : 
$$[\beta_0 | w_c]\sim \mathcal{N}(\textbf{1}^T\tilde Qw_c/\textbf{1}^T\tilde Q\textbf{1}, 1/\textbf{1}^T\tilde Q\textbf{1}),  [w_c|\beta_0]\sim\mathcal{N}( \beta_0 + (\tilde Q + I_n/\tau^2) ^{-1}(z-\beta_0)/\tau^2,(\tilde Q + I_n/\tau^2) ^{-1})$$
The mean of $w_c$ behaves like the mean of $w_s$ except for the term that depends on $\beta_0$ : 
\begin{equation}
\label{eq:fullcond_mean_centered}    
\begin{array}{ccl}
[\overline{w_c}|\beta_0] & \sim & \underbrace{\textbf{1}^T (\tau^2\tilde Q + I_n) ^{-1}(z)/n}_{\text{fixed}}  -
\underbrace{(1-(\textbf{1}^T (\tau^2\tilde Q + I_n) ^{-1}\textbf{1}/n) )\beta_0}_{\text{carry-over}} \\
&&+\underbrace{\mathcal{N}(0,\textbf{1}^T(\tilde Q + I_n/\tau^2)^{-1} \textbf{1}/n^2)}_{\text{innovation}}.
\end{array}
\end{equation}
Injecting the full conditional of $w_c$ into the full conditional of $\beta_0$, we have
$$
\begin{array}{ccl}
[\beta_0^{t+1}|\beta_0^t] &\sim &
\underbrace{\textbf{1}^T \tilde Q (\tilde Q + I_n/\tau^2) ^{-1}z/\tau^2\textbf{1}^T\tilde Q\textbf{1}}_{\text{fixed}}+
\underbrace{\textbf{1}^T \tilde Q (I_n - (\tau^2\tilde Q + I_n) ^{-1})\textbf{1} \beta_0 /\textbf{1}^T\tilde Q\textbf{1}}_{\text{carry-over}}\\
     & &
+ \underbrace{\mathcal{N}(0, \textbf{1}^T\tilde Q(\tilde Q + I_n/\tau^2)^{-1}\tilde Q \textbf{1}/(\textbf{1}^T\tilde Q\textbf{1})^2+1/\textbf{1}^T\tilde Q\textbf{1})}_{\text{innovation}}
\end{array}
$$ 

\underline{Passing to the SVD}\\
Let's compare first the expressions of $[\overline{w_s}|\beta_0]$ and $[\overline{w_c}|\beta_0]$. 
Denote the diagonalization $\tilde Q = V^T \lambda V$, $V$ being a square matrix of eigenvectors and $\lambda$ being a diagonal matrix of eigenvalues. The eigenvalues are positive since $\tilde Q = \tilde R^T\tilde R$. 
Using the fact that adding $I_n$ adds $1$ to every eigenvalue without affecting the eigenvectors, 
$$(\tau^2\tilde Q + I_n) ^{-1}/n = V^T(\tau^2\lambda + I_n)^{-1} V/n.$$
Let $\alpha = (\alpha_1, \ldots, \alpha_n)$ be the coordinates of $\textbf{1}$ in the orthonormal basis defined by $V$. 
$$\textbf{1}^T(\tau^2\tilde Q + I_n) ^{-1}\textbf{1}/n = \Sigma_{i=1}^n\alpha_i^2(\tau^2\lambda + I_n)^{-1}_{i, i}/n.$$ 
Using that $\tilde Q$ is positive-definite on the left and that $\Sigma_1^n\alpha_i = <\textbf{1},\textbf{1}> = n$ on the right, we have
$$
0\leq \textbf{1}^T(\tau^2\tilde Q + I_n) ^{-1}\textbf{1}/n \leq 1
$$ 

As for the centered model, we re-write : 
$$\tilde Q (I_n - (\tau^2\tilde Q + I_n) ^{-1}) = V^T(\lambda (I_n - (\tau^2\lambda + I_n)^{-1})) V = V^T(\tau^2\lambda^2 (\tau^2\lambda + I_n)^{-1}) V.$$
Once this is done, we can express the fraction of $\beta_0^t$ which is conserved in $\beta^{t+1}$ in the centered model as 
$$\textbf{1}^T
\tilde Q (I_n - (\tau^2\tilde Q + I_n) ^{-1}) \textbf{1}/\textbf{1}^T\tilde Q \textbf{1} = \Sigma_{i=1}^n\left((\alpha_i^2\lambda_{i, i})
(\tau^2\lambda_{i, i})/(\tau^2\lambda_{i,i}+1) \right)/\Sigma_{i=1}^n\alpha_i^2\lambda_{i, i}.
$$ 
Like before, thanks to the fact that the eigenvalues of $\tilde Q$ are positive, 
$$0\leq (\tau^2\lambda_{i, i})/(\tau^2\lambda_{i,i}+1)\leq 1$$.

\section{Coloring}
\subsection{Details about the coloring algorithms}
\label{subsection:coloring_algo}

\begin{algorithm}[H]
\begin{algorithmic}
\caption{Naive greedy coloring}
\label{naive_greedy}
\State \textbf{input} $A$\Comment Input adjacency matrix
\State  $(c_1 ,\ldots,  c_n) = (0 ,\ldots, 0)$   \Comment Initialize  colors
\For{$c_i \in (c_1 ,\ldots,  c_n)$}  \Comment Coloration loop
\State $c_i = min ((1 ,\ldots, n)\backslash (c_J)) \text{ with } J = \{J/A_{i, j} = 1\}  $ \Comment{Using smallest available color}
\EndFor
\vspace{5pt}
\State \textbf{return} $(c_1 ,\ldots,  c_n)$
\end{algorithmic}
\end{algorithm}

\begin{algorithm}[H]
\caption{Degree greedy coloring}
\label{degree_greedy}
\begin{algorithmic}
\State \textbf{input} $A$\Comment Input adjacency matrix
\State  $(c_1 ,\ldots,  c_n) = (0 ,\ldots, 0)$   \Comment Initialize  colors
\State  find $(nd_1 ,\ldots, nd_n) = (1 ,\ldots, 1) \cdot A$ \Comment Compute connection degrees of nodes 
\State find  $o(1),\ldots,o(n)$ such that $o(1),\ldots,o(n)$ is a permutation of $1,\ldots,n$  and $i<j \Rightarrow nd_{o(i)}\leq nd_{o(j)}$
\Comment{order nodes by decreasing connection degree}  

 \For{$c_i \in (c_{o(1)} ,\ldots,  c_{o(n)})$}  \Comment Coloration loop
\State $c_i = min ((1 ,\ldots, n)\backslash (c_J)) \text{ with } J = \{J/A_{i, j} = 1\}  $ \Comment{Using smallest available color}
\EndFor
\State \textbf{return} $(c_1 ,\ldots,  c_n)$
\end{algorithmic}
\end{algorithm}

\begin{algorithm}[H]
\caption{DSATUR}
\label{DSATUR}
\begin{algorithmic}
\State \textbf{input} $A$\Comment Input adjacency matrix
\State  $(c_1 ,\ldots,  c_n) = (0 ,\ldots, 0)$   \Comment Initialize  colors
\State  $(sd_1 ,\ldots,  sd_n) = (0 ,\ldots, 0)$  \Comment Initialize saturation degrees
\State  $(nd_1 ,\ldots, nd_n) = (1 ,\ldots, 1) \cdot A$ \Comment Compute connection degrees of nodes 
\While{$0 \in (c_1 ,\ldots,  c_n)$}  \Comment Coloration loop
\State  $j = \{i/c_i =0\}$
\State $j = \{i\in j/sd_i = max_{i\in j}(sd_i)\}$ \Comment{Saturation degree selection rule}
\If{$|j|>1$}
\State $j = \{i\in j/nd_i = max_{i\in j}(nd_i)\}$ \Comment{Node degree tiebreaking rule}
\EndIf
\If{$|j|>1$}
\State Reduce $j$ to its first element \Comment{lexicographical tiebreaking rule}
\EndIf
\State $c_j = min ((1 ,\ldots, n)\backslash (c_{i/A_{i,j}=1}))$ \Comment{Using smallest available color}
\State $sd_{i/A_{i,j}=1} = sd_{i/A_{i,j}=1} +1$  \Comment{Updating saturation degrees}
\EndWhile
\State \textbf{return} $(c_1 ,\ldots,  c_n)$
\end{algorithmic}
\end{algorithm}
\clearpage
\subsection{Results of coloring experiments}
\label{subsection:coloring_exp}

\begin{table*}[!htbp]
    \caption{Case-by-case mean number of colors in the pilot experiment.}
    \label{tab:table_means_exp1}
    \centering
    \begin{tabular}{llllllllllll}
    & & & \multicolumn{3}{c}{Coordinate ordering} & \multicolumn{3}{c}{Max-min ordering} & \multicolumn{3}{c}{Random ordering} \\
    m & n & d & degree & dsatur & naive & degree & dsatur & naive& degree & dsatur & naive\\
    \hline
\multirow{6}{*}{5}  & \multirow{2}{*}{500}  & 2 &   7.4 &   6.0 &  6.0 &  9.8 &  9.1 & 10.1 & 10.3 &   9.2 & 10.0\\
                    &                       & 3 &   7.9 &   6.0 &  6.0 & 10.8 &  9.6 & 11.0 & 10.6 &  10.0 & 11.1\\
                    & \multirow{2}{*}{1000} & 2 &   8.2 &   6.0 &  6.0 & 10.1 &  9.4 & 10.3 & 10.6 &   9.3 & 10.2\\
                    &                       & 3 &   8.0 &   6.0 &  6.0 & 11.8 & 10.0 & 11.1 & 11.3 &  10.1 & 11.1\\
                    & \multirow{2}{*}{2000} & 2 &   8.6 &   6.0 &  6.0 & 10.5 &  9.6 & 10.3 & 10.7 &   9.9 & 10.2\\
                    &                       & 3 &   8.9 &   6.0 &  6.0 & 11.6 & 10.1 & 11.7 & 11.8 &  10.0 & 11.4\\
    \hdashline
\multirow{6}{*}{10} & \multirow{2}{*}{500}  & 2 &  12.9 &  11.0 & 11.0 & 18.7 & 17.4 & 18.9 & 19.0 &  17.8 & 19.4\\
                    &                       & 3 &  13.0 &  11.0 & 11.0 & 20.9 & 19.1 & 21.2 & 21.0 &  18.9 & 21.3\\
                    & \multirow{2}{*}{1000} & 2 &  13.7 &  11.0 & 11.0 & 19.2 & 17.8 & 19.5 & 19.9 &  17.9 & 20.0\\
                    &                       & 3 &  13.6 &  11.0 & 11.0 & 21.3 & 19.4 & 22.3 & 21.6 &  19.4 & 22.0\\
                    & \multirow{2}{*}{2000} & 2 &  14.8 &  11.0 & 11.0 & 20.2 & 18.2 & 20.0 & 20.7 &  18.6 & 19.8\\
                    &                       & 3 &  15.0 &  11.0 & 11.0 & 22.3 & 19.7 & 22.6 & 22.7 &  19.9 & 22.4\\
    \hdashline
\multirow{6}{*}{20} & \multirow{2}{*}{500}  & 2 &  23.0 &  21.0 & 21.0 & 36.5 & 33.9 & 37.4 & 36.8 &  34.2 & 37.5\\
                    &                       & 3 &  23.3 &  21.0 & 21.0 & 40.8 & 37.1 & 44.1 & 40.5 &  37.5 & 43.4\\
                    & \multirow{2}{*}{1000} & 2 &  24.9 &  21.0 & 21.0 & 37.6 & 35.0 & 38.2 & 38.6 &  35.2 & 38.5\\
                    &                       & 3 &  23.9 &  21.0 & 21.0 & 42.9 & 38.4 & 45.3 & 43.1 &  38.8 & 44.7\\
                    & \multirow{2}{*}{2000} & 2 &  26.0 &  21.0 & 21.0 & 38.6 & 35.9 & 38.8 & 39.1 &  36.2 & 39.1\\
                    &                       & 3 &  25.7 &  21.0 & 21.0 & 44.8 & 39.6 & 46.4 & 44.7 &  40.0 & 45.6\\
    \end{tabular}
\end{table*}

\begin{table*}[]
    \caption{Case-by-case mean number of colors for large graphs.}
    \label{tab:table_means_exp2}
    \centering
    \begin{tabular}{lllllllll}
    & & & \multicolumn{2}{c}{Coordinate ordering} & \multicolumn{2}{c}{Max-min ordering} & \multicolumn{2}{c}{Random ordering} \\
    m & n & d & degree & naive & degree  & naive& degree & naive\\
    \hline
\multirow{6}{*}{5}  & \multirow{2}{*}{50000}  & 2 &  10.0   &   8.8  &  11.3  & 11.0  &  12.3  &  11.1 \\
                    &                         & 3 &  10.0   &   8.7  &  13.1  & 12.6  &  13.1  &  12.3 \\
                    & \multirow{2}{*}{100000} & 2 &  10.1   &   9.0  &  11.7  & 11.0  &  12.1  &  11.0 \\
                    &                         & 3 &  10.0   &   9.1  &  13.1  & 13.0  &  13.2  &  12.3 \\
                    & \multirow{2}{*}{200000} & 2 &  10.1   &   9.3  &  11.9  & 11.2  &  12.1  &  11.4 \\
                    &                         & 3 &  10.3   &   9.7  &  13.3  & 13.0  &  13.5  &  12.8 \\
    \hdashline
\multirow{6}{*}{10} & \multirow{2}{*}{50000}  & 2 &  17.1   &  13.8  &  21.5  & 21.0  &  22.7  &  20.8 \\
                    &                         & 3 &  17.0   &  14.0  &  24.5  & 24.1  &  25.5  &  23.7 \\
                    & \multirow{2}{*}{100000} & 2 &  17.0   &  15.6  &  22.0  & 21.2  &  22.9  &  21.0 \\
                    &                         & 3 &  17.0   &  15.5  &  24.9  & 24.2  &  25.6  &  23.9 \\
                    & \multirow{2}{*}{200000} & 2 &  17.9   &  16.7  &  22.2  & 21.3  &  22.9  &  21.1 \\
                    &                         & 3 &  18.0   &  16.8  &  25.2  & 24.2  &  26.4  &  24.3 \\
    \hdashline
\multirow{6}{*}{20} & \multirow{2}{*}{50000}  & 2 &  31.4   &  21.1  &  41.4  & 40.1  &  43.0  &  40.4 \\
                    &                         & 3 &  31.2   &  21.6  &  49.7  & 48.2  &  50.2  &  47.8 \\
                    & \multirow{2}{*}{100000} & 2 &  30.8   &  25.1  &  41.9  & 40.7  &  44.3  &  40.8 \\
                    &                         & 3 &  31.0   &  24.6  &  50.0  & 48.5  &  50.8  &  47.6 \\
                    & \multirow{2}{*}{200000} & 2 &  30.3   &  28.6  &  41.7  & 40.7  &  44.3  &  40.8 \\
                    &                         & 3 &  30.2   &  28.7  &  50.5  & 48.9  &  51.6  &  48.2 \\
    \end{tabular}
\end{table*}

\begin{table*}[]
    \caption{Case-by-case mean number of colors for blocked graphs.}
    \label{tab:table_means_exp3}
    \centering
    \begin{tabular}{llllllllllll}
    & & & \multicolumn{3}{c}{Coordinate ordering} & \multicolumn{3}{c}{Max-min ordering} & \multicolumn{3}{c}{Random ordering} \\
    m & blocks & d & degree & dsatur & naive & degree & dsatur & naive& degree & dsatur & naive\\
    \hline
\multirow{10}{*}{5} & \multirow{2}{*}{10} & 2 & 2.9 & 2.5 & 3.0 &  6.2 &  6.2 &  6.6 &  6.5 &  6.5 & 6.7    \\  
                    &                     & 3 & 2.7 & 2.3 & 2.7 &  7.5 &  7.5 &  7.5 &  7.3 &  7.3 & 7.5    \\   
                    & \multirow{2}{*}{20} & 2 & 3.0 & 2.4 & 3.0 &  7.9 &  7.9 &  8.6 &  7.6 &  7.5 & 8.3    \\ 
                    &                     & 3 & 3.0 & 2.6 & 3.0 &  9.0 &  8.5 &  9.6 &  8.6 &  8.3 & 9.5    \\   
                    & \multirow{2}{*}{50} & 2 & 3.0 & 2.4 & 3.0 &  8.9 &  8.6 & 10.8 &  8.6 &  8.1 & 10.2   \\  
                    &                     & 3 & 3.0 & 2.4 & 3.0 & 10.9 & 10.2 & 12.7 & 10.3 &  9.4 & 12.0   \\    
                    & \multirow{2}{*}{100}& 2 & 3.0 & 2.4 & 3.0 &  9.8 &  9.1 & 12.0 &  9.0 &  8.6 & 11.3   \\   
                    &                     & 3 & 3.0 & 2.2 & 3.0 & 11.8 & 11.1 & 14.0 & 11.0 & 10.2 & 13.0   \\    
                    & \multirow{2}{*}{500}& 2 & 3.0 & 2.9 & 3.1 & 10.3 &  9.3 & 14.7 & 10.1 &  9.1 & 14.3   \\   
                    &                     & 3 & 3.0 & 3.0 & 3.0 & 13.2 & 11.7 & 17.3 & 12.4 & 11.1 & 16.1   \\    
    \hdashline
\multirow{10}{*}{10}& \multirow{2}{*}{10} & 2 & 2.9 & 2.5 & 3.0 &  8.1 &  8.1 &  8.2 &  8.6 &  8.6 & 8.6    \\ 
                    &                     & 3 & 2.7 & 2.3 & 2.7 &  9.3 &  9.3 &  9.3 &  9.3 &  9.3 & 9.3    \\ 
                    & \multirow{2}{*}{20} & 2 & 3.0 & 2.4 & 3.0 & 11.4 & 11.4 & 12.2 & 11.9 & 11.9 & 12.5   \\  
                    &                     & 3 & 3.0 & 2.6 & 3.0 & 13.5 & 13.4 & 13.8 & 12.5 & 12.3 & 13.0   \\  
                    & \multirow{2}{*}{50} & 2 & 3.0 & 2.4 & 3.0 & 14.8 & 14.2 & 17.1 & 14.0 & 13.6 & 16.0   \\  
                    &                     & 3 & 3.0 & 2.4 & 3.0 & 17.5 & 16.6 & 19.0 & 16.2 & 15.5 & 18.2   \\  
                    & \multirow{2}{*}{100}& 2 & 3.0 & 2.4 & 3.0 & 16.7 & 16.0 & 20.6 & 15.7 & 15.0 & 19.2   \\  
                    &                     & 3 & 3.0 & 2.2 & 3.0 & 19.8 & 18.5 & 22.9 & 18.0 & 17.1 & 21.2   \\  
                    & \multirow{2}{*}{500}& 2 & 3.6 & 3.1 & 4.1 & 19.6 & 18.2 & 26.5 & 18.6 & 17.0 & 24.5   \\  
                    &                     & 3 & 3.8 & 3.4 & 4.3 & 22.9 & 21.0 & 30.9 & 20.9 & 18.8 & 28.0   \\  
    \hdashline
\multirow{10}{*}{20}& \multirow{2}{*}{10} & 2 & 2.9 & 2.5 & 3.0 &  9.8 &  9.8 &  9.8 &  9.9 &  9.9 & 9.9    \\ 
                    &                     & 3 & 2.7 & 2.3 & 2.7 & 10.0 & 10.0 & 10.0 & 10.0 & 10.0 & 10.0   \\  
                    & \multirow{2}{*}{20} & 2 & 3.0 & 2.4 & 3.0 & 16.2 & 16.2 & 16.7 & 16.3 & 16.3 & 16.5   \\  
                    &                     & 3 & 3.0 & 2.6 & 3.0 & 17.7 & 17.7 & 17.7 & 17.1 & 17.1 & 17.2   \\  
                    & \multirow{2}{*}{50} & 2 & 3.0 & 2.4 & 3.0 & 25.1 & 24.5 & 27.0 & 23.0 & 22.9 & 25.4   \\  
                    &                     & 3 & 3.0 & 2.4 & 3.0 & 27.9 & 27.8 & 30.4 & 24.8 & 24.3 & 27.4   \\  
                    & \multirow{2}{*}{100}& 2 & 3.0 & 2.4 & 3.0 & 30.4 & 29.1 & 34.6 & 26.7 & 25.7 & 31.0   \\  
                    &                     & 3 & 3.0 & 2.2 & 3.0 & 33.7 & 32.4 & 38.2 & 29.8 & 28.9 & 33.9   \\  
                    & \multirow{2}{*}{500}& 2 & 5.2 & 4.7 & 5.6 & 37.2 & 35.2 & 47.6 & 34.0 & 31.2 & 42.6   \\  
                    &                     & 3 & 5.2 & 4.8 & 5.8 & 42.0 & 39.0 & 55.0 & 38.9 & 35.1 & 49.1   \\                        
    \end{tabular}
\end{table*}

\end{document}